\documentclass[10pt, two column, twoside]{IEEEtran}

\usepackage{amssymb}
\usepackage{amsmath}
\usepackage{mathrsfs}

\usepackage{pgfplots}
\usepackage{tikz}
\usetikzlibrary{arrows}
\usepackage{subfigure}
\usepackage{graphicx,booktabs,multirow}

\definecolor{colorhkust}{RGB}{20,43,140}
\definecolor{colortsinghua}{RGB}{116,52,129}
\definecolor{color1}{HTML}{D0B22B}

\newcommand{\bs}{\boldsymbol}
\newcommand{\rev}{\color{black}}

\newtheorem{proposition}{Proposition}
\newtheorem{remark}{Remark}

\begin{document}

\title{Large-Scale Convex Optimization for Dense Wireless Cooperative Networks}
\author{Yuanming~Shi,
        Jun~Zhang,
        Brendan O'Donoghue,
        and~Khaled~B.~Letaief,~\IEEEmembership{Fellow,~IEEE}
\thanks{Copyright (c) 2015 IEEE. Personal use of this material is permitted. However, permission to use this material for any other purposes must be obtained from the IEEE by sending a request to pubs-permissions@ieee.org. This work
is supported by the Hong Kong Research Grant Council under Grant No. 16200214.}
\thanks{Y. Shi, J. Zhang and K. B. Letaief are with the Department of Electronic and Computer Engineering, Hong Kong University of Science and Technology (e-mail: \{yshiac, eejzhang, eekhaled\}@ust.hk).}
\thanks{B. O'Donoghue is with the Electrical Engineering Department, Stanford University, Stanford, CA 94305 USA (e-mail: bodono@stanford.edu).}}

\maketitle

\begin{abstract}
Convex optimization  is a powerful tool for resource allocation and signal processing in wireless networks. As the network density is expected to drastically increase in order to accommodate the exponentially growing mobile data traffic,  performance optimization problems are entering a new era characterized by a high dimension and/or a large number of constraints, which poses significant design and computational challenges. In this paper, we present a novel two-stage approach to solve large-scale convex optimization problems for dense wireless cooperative networks, which can effectively detect infeasibility and enjoy modeling flexibility. In the proposed approach, the original large-scale  convex problem is transformed into a standard cone programming form in the first stage  via matrix stuffing, which only needs to copy the problem parameters such as  channel state information (CSI) and
quality-of-service (QoS) requirements to the pre-stored structure of the standard form. The capability of yielding infeasibility certificates  and enabling parallel
computing is achieved by solving the homogeneous self-dual embedding of the
primal-dual pair of the standard form. In the solving stage, the operator splitting method, namely, the alternating direction method of multipliers (ADMM), is adopted to solve the large-scale homogeneous self-dual embedding. Compared with second-order methods, ADMM  can solve large-scale problems in parallel with modest accuracy within a reasonable amount of time.  Simulation results will demonstrate the speedup, scalability,
and reliability of the proposed framework compared with the state-of-the-art modeling frameworks and
solvers.   
\end{abstract}


\section{Introduction}
\IEEEPARstart{T}{he} proliferation of smart mobile devices, coupled with new types of wireless applications, has led to an exponential growth of wireless and mobile data traffic. In order to provide high-volume and diversified data services, ultra-dense wireless cooperative network architectures have been proposed for next generation wireless networks \cite{Yuanming_WCMLargeCVX}, e.g., Cloud-RAN \cite{mobile2011c, Yuanming_TWC2014}, and distributed antenna systems \cite{zhou2003distributed}. To enable efficient interference management and resource allocation, large-scale multi-entity collaboration will play pivotal roles in dense wireless networks. For instance, in Cloud-RAN, all the baseband signal processing is shifted to a single cloud data center  with very powerful computational capability. Thus the centralized signal processing can be performed to support large-scale cooperative transmission/reception among the radio access units (RAUs).

Convex optimization serves as an indispensable tool for   resource allocation and signal processing in wireless communication systems \cite{gershman2010convex,Z.Q.Luo_SPM2010, Bjornson_TCIT2013}. For instance, coordinated beamforming  \cite{WeiYu_WC10} often yields a direct convex optimization formulation, i.e., second-order cone programming (SOCP) \cite{boyd2004convex}. The network max-min fairness rate optimization \cite{Hanly_2012TIT} can be solved through the bi-section method \cite{boyd2004convex} in polynomial time, wherein a sequence of convex subproblems
are  solved. Furthermore, convex relaxation provides a principled way of developing polynomial-time algorithms for non-convex or NP-hard problems, e.g., group-sparsity penalty relaxation for the NP-hard mixed integer nonlinear programming problems \cite{Yuanming_TWC2014}, semidefinite relaxation \cite{Z.Q.Luo_SPM2010} for NP-hard robust beamforming  \cite{shen2012distributed,Yuanming_SDP2014} and multicast beamforming \cite{Yuanming_ICC2015SDP}, and sequential convex approximation to the highly intractable stochastic coordinated beamforming \cite{Yuanming_TSP14SCB}.

Nevertheless, in dense wireless cooperative networks \cite{Yuanming_WCMLargeCVX}, which may possibly need to simultaneously handle hundreds of RAUs, resource
allocation and signal processing problems will be dramatically scaled up. The underlying optimization problems will have  high dimensions and/or large numbers of constraints (e.g., per-RAU transmit power constraints and per-MU (mobile user) QoS constraints). For instance, for a Cloud-RAN with 100 single-antenna RAUs and 100 single-antenna MUs, the dimension of the aggregative coordinated beamforming vector (i.e., the optimization variables) will be $10^{4}$.  Most advanced off-the-shelf solvers  (e.g., SeDuMi \cite{SeDuMi_1999using}, SDPT3 \cite{SDPT3_1999} and MOSEK \cite{Mosek_2000mosek})  are based on the interior-point method. However, the computational burden of such second-order method makes it inapplicable for  large-scale problems. For instance, solving convex quadratic programs has cubic complexity \cite{nesterov1994interior}. Furthermore, to use these solvers, the original problems need to be transformed to  the standard forms supported by the solvers. Although the parser/solver modeling frameworks like CVX \cite{cvx} and YALMIP \cite{YALMIP_2004yalmip} can automatically transform the original problem instances into standard forms, it may require substantial time to perform such transformation \cite{boyd_code2013}, especially for problems with a large number of constraints \cite{Yuanming_Globecom2014}.  

One may also develop custom algorithms to enable efficient computation by exploiting the structures of specific problems. For instance, the uplink-downlink duality \cite{WeiYu_WC10} is exploited to extract the structures of the optimal beamformers \cite{Bjornson_SPM2014} and enable efficient algorithms. However, such an approach still has the cubic complexity to perform matrix inversion at each iteration \cite{Luo_2013base}. First-order methods, e.g., the ADMM algorithm \cite{boyd2011distributed}, have recently attracted attention for their distributed  and parallelizable implementation, as well as the capability of scaling to large problem sizes. However, most existing ADMM based algorithms cannot provide the certificates of infeasibility \cite{shen2012distributed, Luo_2013base,joshi2014distributed}. Furthermore, some of them may still fail to scale to large problem sizes, due to the SOCP subproblems \cite{joshi2014distributed} or semidefinite programming (SDP) subproblems \cite{shen2012distributed} needed to be solved at each iteration.

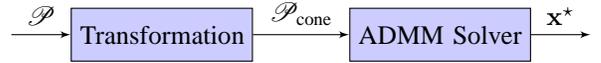
\begin{figure}[!t]
\centering
\tikzstyle{int}=[draw, fill=blue!20, minimum size=2em]
\tikzstyle{init} = [pin edge={to-,thin,black}]
\begin{tikzpicture}
    [node distance=3.7cm,auto,>=latex']
    \node [int, ] (a) {Transformation};
    \node (b) [left of=a,node distance=2cm, coordinate] {a};
    \node [int, ] (c) [right of=a] {ADMM Solver};
    \node [coordinate] (end) [right of=c, node distance=2.0cm]{};
    \path[->] (b) edge node {$\mathscr{P}$} (a);
    \path[->] (a) edge node {$\mathscr{P}_{\textrm{cone}}$} (c);
    \draw[->] (c) edge node {$\bf{x}^{\star}$} (end) ;
\end{tikzpicture}
\caption{The proposed two-stage approach for  large-scale convex optimization. The optimal solution or the certificate\ of infeasibility can be extracted from ${\bf{x}}^{\star}$ by the ADMM solver.}
\label{LSCO}
\vspace*{-15pt}
\end{figure}

Without efficient and scalable algorithms, previous studies of wireless cooperative networks either only demonstrate performance in small-size networks, typically with less than 10 RAUs, or resort to sub-optimal algorithms, e.g., zero-forcing based approaches \cite{Jun_2009networkedTWC,Goldsmith_JSAC2006optimality}. Meanwhile, from the above discussion, we see that the large-scale optimization algorithms to be developed should possess the following two features:
\begin{itemize}
\item To scale well to large problem sizes with parallel computing capability;
\item To effectively detect problem infeasibility, i.e., provide certificates of infeasibility.
\end{itemize} 
To address these two challenges in a unified way, in this paper, we shall propose a two-stage approach as shown in Fig. {\ref{LSCO}}. The proposed framework is capable to solve large-scale convex optimization problems in parallel, as well as providing certificates of infeasibility. Specifically, the original problem $\mathscr{P}$ will be first transformed into a standard cone programming form $\mathscr{P}_{\textrm{cone}}$ \cite{nesterov1994interior} based on the Smith form reformulation \cite{smith1996optimal}, via introducing a new variable for each subexpression in the disciplined convex programming form \cite{boyd2008graph} of the original problem. This will eventually transform the coupled constraints in the original problem into the constraint only consisting of two convex sets: a subspace and a convex set formed by a Cartesian product of a finite number of standard convex cones. Such a structure helps to develop efficient parallelizable algorithms and enable the infeasibility
detection capability \emph{simultaneously} via solving the homogeneous self-dual embedding \cite{ye1994nl}
of the primal-dual pair of the standard form by the ADMM algorithm. 

As the mapping between the standard cone program and the original problem only depends on the network size (i.e., the numbers of RAUs, MUs and antennas at each RAU), we can pre-generate and store the structures of the standard forms with different candidate network sizes.  Then for each problem instance, that is, given the channel coefficients, QoS requirements, and maximum RAU transmit powers, we only need to copy the original problem parameters to the standard cone programming data. Thus, the transformation procedure can be very efficient and can avoid repeatedly parsing and re-generating problems \cite{cvx,YALMIP_2004yalmip}. This technique is called \emph{matrix stuffing} \cite{boyd_code2013,Yuanming_Globecom2014}, which is essential for the proposed  framework to scale well to large problem sizes. It may also help rapid prototyping and testing for practical equipment development.

\subsection{Contributions}

The major contributions of the paper are summarized as follows: 
\begin{enumerate}
\item We formulate main performance optimization problems  in dense wireless cooperative networks into a general framework. It is shown that all of them can essentially be solved through solving one or a sequence of large-scale convex optimization or convex feasibility problems. 
\item To enable both the infeasibility detection capability and parallel computing capability, we propose to transform the original convex problem to an equivalent standard cone program. The transformation procedure scales very well to large problem sizes with the matrix stuffing technique.  Simulation results will demonstrate the effectiveness of the proposed fast transformation approach over the state-of-art parser/solver modeling frameworks. 

\item The operator splitting method is then adopted to solve the large-scale homogeneous self-dual embedding of the primal-dual pair of the transformed standard cone program in parallel. This first-order optimization algorithm makes the second stage scalable. Simulation results will show that it  can speedup several orders of magnitude over the state-of-art interior-point solvers. 

\item The proposed framework enables evaluating various cooperation strategies in dense wireless networks, and helps reveal new insights {\rev{\emph{numerically}}}. {\rev{For instance}}, simulation results demonstrate a significant performance gain of optimal beamforming over sub-optimal schemes, which shows the importance of developing large-scale optimal beamforming algorithms. \end{enumerate} 

This work will serve the purpose of providing  practical and theoretical guidelines  on designing algorithms for generic large-scale optimization problems in dense wireless networks.

\subsection{Organization}
The remainder of the paper is organized as follows. Section {\ref{probf}} presents the
system model and problem formulations.
In Section {\ref{hsde}}, a systematic cone programming form transformation procedure is developed.  The operator splitting method is presented in Section {\ref{opsp}}. The practical implementation issues are discussed in Section \ref{pii}. Numerical results will be demonstrated in Section {\ref{num}}. Finally, conclusions and discussions are presented in Section {\ref{concl}}. To keep the main
text clean and free of technical details, we divert most of the proofs, derivations to
the appendix. 

\section{Large-Scale Optimization in Dense wireless Cooperative Networks}
\label{probf}
In this section, we will first present two representative optimization problems in wireless cooperative networks, i.e., the network power minimization problem and the network utility maximization problem. We will then provide a unified formulation for large-scale optimization problems in dense wireless cooperative networks.
\begin{figure}[t]
  \centering
  \includegraphics[width=0.95\columnwidth]{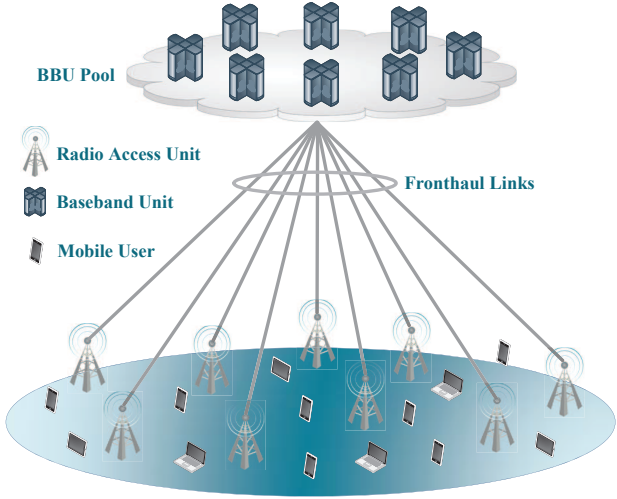}
 \caption{\rev{The architecture of Cloud-RAN, in which, all the RAUs  are connected to a BBU pool through high-capacity and low-latency optical fronthaul links. To enable full cooperation among RAUs, it is assumed that all the user data and CSI are available at the BBU pool.}}
 \label{cran}
 \end{figure}

\subsection{Signal Model}
Consider a dense fully cooperative network{\footnote{\rev{The full cooperation among all the RAUs with global CSI and full user data sharing is used as an illustrative example. The proposed framework can be  extended to more general cooperation scenarios, e.g., with partial user data sharing among RAUs as presented in \cite[Section 1.3.1]{Bjornson_TCIT2013}.}}} with $L$ RAUs and $K$ single-antenna MUs, where the $l$-th RAU is equipped with $N_{l}$ antennas.  The centralized signal processing is performed at a central processor, e.g., the baseband unit  pool in Cloud-RAN \cite{mobile2011c,Yuanming_TWC2014} as shown in Fig. {\ref{cran}}. The propagation channel from the $l$-th RAU to the $k$-th MU is denoted as ${\bf{h}}_{kl}\in\mathbb{C}^{N_{l}}, \forall k, l$. {\rev{In this paper, we focus on the downlink transmission for illustrative purpose. But our proposed approach can also be applied in the uplink transmission, as we only need to exploit the convexity of the resulting performance optimization problems.}} The received signal $y_{k}\in\mathbb{C}$ at MU $k$ is given by
\setlength\arraycolsep{1pt}
\begin{eqnarray}
y_{k}=\sum_{l=1}^{L}{\bf{h}}_{kl}^{\sf{H}}{\bf{v}}_{lk}s_{k}+\sum_{i\ne k}\sum_{l=1}^{L}{\bf{h}}_{kl}^{\sf{H}}{\bf{v}}_{li}s_{i}+n_{k},
\forall k,
\end{eqnarray}
where $s_{k}$ is the encoded information symbol for MU $k$ with $\mathbb{E}[|s_{k}|^2]=1$,
${\bf{v}}_{lk}\in\mathbb{C}^{N_{l}}$ is the transmit beamforming vector from
the $l$-th RAU to the $k$-th MU, and $n_{k}\sim\mathcal{CN}(0, \sigma_{k}^2)$
is the additive Gaussian noise at MU $k$. We assume that $s_{k}$'s and $n_{k}$'s
are mutually independent and all the users apply single user detection. Thus the
signal-to-interference-plus-noise ratio (SINR) of MU $k$ is
given by
\begin{eqnarray}
{{\Gamma}}_{k}({\bf{v}})={{|{\bf{h}}_{k}^{\sf{H}}{\bf{v}}_{k}|^2}\over{\sum_{i\ne
k}|{\bf{h}}_{k}^{\sf{H}}{\bf{v}}_{i}|^2+\sigma_{k}^2}}, \forall k,
\end{eqnarray}
where ${\bf{h}}_{k}\triangleq [{\bf{h}}_{k1}^{T},\dots, {\bf{h}}_{kL}^{T}]^{T}\in\mathbb{C}^{N}$ with $N=\sum_{l=1}^{L}N_{l}$, ${\bf{v}}_{k}\triangleq[{\bf{v}}_{1k}^{T},
{\bf{v}}_{2k}^{T},\dots, {\bf{v}}_{Lk}^{T}]^{T}\in\mathbb{C}^{N}$ and ${\bf{v}}\triangleq
[{\bf{v}}_1^{T},\dots, {\bf{v}}_{K}^{T}]^{T}\in\mathbb{C}^{NK}$. We assume that each RAU has its own  power constraint,  
\begin{eqnarray}
\sum_{k=1}^{K}\|{\bf{v}}_{lk}\|_2^2\le P_{l}, \forall l,
\end{eqnarray}
where $P_l>0$ is the maximum transmit power of the $l$-th RAU. In this paper, we assume that the full and perfect CSI is available at the central processor and all RAUs only provide unicast/broadcast services. 
 
\subsection{Network Power Minimization}
Network power consumption is an important performance metric for the energy efficiency design in wireless cooperative networks.
Coordinated beamforming is an efficient way to design energy-efficient systems \cite{WeiYu_WC10}, in which, beamforming vectors ${\bf{v}}_{lk}$'s are designed to minimize the total transmit power among RAUs while satisfying the QoS
requirements for all the MUs. Specifically, given the target SINRs ${\bs{\gamma}}=(\gamma_1,\dots, \gamma_K)$ for all the MUs  with $\gamma_k> 0$, $\forall k$, we will solve the following total transmit power minimization problem: 
\begin{eqnarray}
\mathscr{P}_1 ({\boldsymbol{\gamma}}):
\mathop {\rm{minimize~}}_{{\bf{v}}\in\mathcal{V}} && \sum_{l=1}^{L}\sum_{k=1}^{K}\|{\bf{v}}_{lk}\|_2^2,
\end{eqnarray}
where $\mathcal{V}$ is the intersection of the sets formed by transmit power constraints and QoS constraints, i.e., 
\begin{eqnarray}
\label{intset}
\mathcal{V}=\mathcal{P}_{1}\cap\mathcal{P}_2\cap\dots\cap\mathcal{P}_{L}\cap\mathcal{Q}_1\cap\mathcal{Q}_2\dots,\cap\mathcal{Q}_{K},
\end{eqnarray}
where $\mathcal{P}_l$ 's are feasible sets of ${\bf{v}}$ that satisfy the
 per-RAU transmit power constraints, i.e.,
\begin{eqnarray}
\label{pconstraint}
 {\mathcal{P}}_l=\left\{{\bf{v}}\in\mathbb{C}^{NK}: \sum_{k=1}^{K}\|{\bf{v}}_{lk}\|_2^2\le
P_{l}\right\}, \forall l,
\end{eqnarray}
and $\mathcal{Q}_k$'s are the feasible sets of $\bf{v}$ that satisfy the  per-MU QoS constraints, i.e.,
\begin{eqnarray}
{\mathcal{Q}}_{k}=\{{\bf{v}}\in\mathbb{C}^{NK}: {{\Gamma}}_{k}({\bf{v}})\ge\gamma_k\}, \forall k.
\end{eqnarray}
As all the sets ${\mathcal{Q}}_k$'s and $\mathcal{P}_l$'s can be reformulated into second-order cones as shown in \cite{Yuanming_TWC2014}, problem $\mathscr{P}_{1}({\bs{\gamma}})$ can be reformulated as an SOCP problem.

However, in dense wireless cooperative networks, the mobile hauling network consumption can not be ignored. In \cite{Yuanming_TWC2014}, a two-stage group sparse beamforming (GSBF) framework is proposed to minimize the network power consumption for Cloud-RAN, including the power consumption of all  optical fronthaul links and the transmit power consumption of all  RAUs. Specially, in the first stage, the group-sparsity structure of the aggregated beamformer ${\bf{v}}$ is induced by minimizing the weighted mixed $\ell_1/\ell_2$-norm of ${\bf{v}}$, i.e., 
\begin{eqnarray}
\mathscr{P}_2 ({\boldsymbol{\gamma}}):
\mathop {\rm{minimize~}}_{{\bf{v}}\in\mathcal{V}} &&\sum_{l=1}^{L}\omega_{l}\|\tilde{\bf{v}}_{l}\|_2,
\end{eqnarray}
where $\tilde{\bf{v}}_{l}=[{\bf{v}}_{l1}^{T},\dots,
{\bf{v}}_{lK}^{T}]^{T}\in\mathbb{C}^{N_lK}$ is the aggregated beamforming
vector at RAU $l$, and $\omega_l>0$ is the corresponding weight for the beamformer coefficient group $\tilde{\bf{v}}_l$. Based on the (approximated) group sparse beamformer ${\bf{v}}^{\star}$, which is the optimal solution to $\mathscr{P}_2(\bs{\gamma})$, in the second stage, an RAU selection procedure is performed to switch off some RAUs so as to minimize the network power consumption. In this procedure,
we need to check if the remaining RAUs can support the QoS requirements for all the MUs, i.e., check the feasibility of problem $\mathscr{P}_1({\bs{\gamma}})$  given the active RAUs. Please refer to \cite{Yuanming_TWC2014} for more details on the group sparse beamforming algorithm.

\subsection{Network Utility Maximization}
Network utility maximization is a general approach to optimize network performance.
We consider maximizing an arbitrary network utility function $U({{\Gamma}}_1({\bf{v}}), \dots, {{\Gamma}}_K({\bf{v}}))$ that is strictly increasing
in the SINR of each MU \cite{Bjornson_TCIT2013}, i.e., 
\begin{eqnarray}
\mathscr{P}_3:
\mathop {\rm{maximize~}}_{{\bf{v}}\in\mathcal{V}_1}&&U({{\Gamma}}_1({\bf{v}}),\dots,
{{\Gamma}}_K({\bf{v}})),
\end{eqnarray}
where $\mathcal{V}_1=\cap_{l=1}^{L}\mathcal{P}_l$ is the intersection of the sets of the per-RAU transmit power constraints (\ref{pconstraint}). It is generally very difficult to solve, though there are tremendous research
efforts on this problem \cite{Bjornson_TCIT2013}. In particular, Liu {\emph{et al.}} in \cite{Luo_TSP2011coordinated} proved that $\mathscr{P}_3$ is   NP-hard for many common utility functions, e.g., weighted sum-rate. Please refer to \cite[Table 2.1]{Bjornson_TCIT2013} for details on classification of the convexity of utility optimization problems.

Assume that we have the prior knowledge of SINR values $\Gamma_1^{\star},\dots, \Gamma_K^{\star}$ that can be achieved by the optimal solution to problem $\mathscr{P}_{3}$. Then the optimal solution to problem $\mathscr{P}_1(\bs{\gamma})$ with target SINRs as ${\bs{\gamma}}=(\Gamma_1^{\star},\dots, \Gamma_K^{\star})$ is an optimal solution to problem $\mathscr{P}_3$ as well \cite{Bjornson_SPM2014}. The difference between  problem $\mathscr{P}_1(\bs{\gamma})$  and problem $\mathscr{P}_3$ is that the SINRs in $\mathscr{P}_1(\bs{\gamma})$  are pre-defined, while the optimal SINRs in $\mathscr{P}_3$ need to be searched.  For the max-min fairness maximization problem, optimal SINRs can be searched by the bi-section method \cite{Yuanming_Globecom2014}, which can be accomplished in polynomial time. For the general increasing utility  maximization problem $\mathscr{P}_3$, the corresponding optimal SINRs can be searched as follows
\begin{eqnarray}
\label{mon}
\mathop {\rm{maximize~}}_{{\bs{\gamma}}\in\mathcal{R}}&&U(\gamma_1,\dots,
\gamma_K),
\end{eqnarray}
where $\mathcal{R}\in\mathbb{R}_{+}^{K}$ is the achievable performance region \begin{eqnarray}
\mathcal{R}=\{({{\Gamma}}_1({\bf{v}}), \dots, {{\Gamma}}_K({\bf{v}})): {\bf{v}}\in\mathcal{V}_1\}.
\end{eqnarray}
Problem (\ref{mon}) is a  monotonic
optimization problem \cite{Tuy_2000monotonic} and thus can be solved by the polyblock outer approximation algorithm \cite{Tuy_2000monotonic} or the branch-reduce-and-bound algorithm \cite{Bjornson_TCIT2013}. The general idea of both algorithms is iteratively improving the lower-bound $U_{\textrm{min}}$ and upper-bound $U_{\textrm{max}}$ of the objective function of problem (\ref{mon}) such that 
\begin{eqnarray}
U_{\textrm{max}}-U_{\textrm{min}}\le\epsilon,
\end{eqnarray}
for a given accuracy $\epsilon$ in finite iterations. In particular, at the $m$-iteration, we need to check the convex feasibility problem of $\mathscr{P}_1(\bs{\gamma}^{[m]})$ given the target SINRs ${\bs{\gamma}}^{[m]}=(\Gamma_{1}^{[m]},\dots, \Gamma_K^{[m]})$. However, the number
of iterations scales exponentially with the number of MUs \cite{Bjornson_TCIT2013}. Please refer to the tutorial \cite[Section 2.3]{Bjornson_TCIT2013} for more details. Furthermore, the network achievable rate region \cite{Jorswieck_TSP2008complete} can also be characterized by the rate profile method \cite{zhang2010cooperative} via solving a sequence of such convex feasibility problems $\mathscr{P}_1(\bs{\gamma})$.

\subsection{A Unified Framework of Large-Scale Network Optimization}
In dense wireless cooperative networks, the central processor can support hundreds of RAUs for simultaneously transmission/reception \cite{mobile2011c}. Therefore, all the above optimization problems are shifted into a new domain with a high problem dimension and   a large number of constraints. As presented previously, to solve the performance optimization problems, we essentially need to solve a sequence of the following convex optimization problem with different problem instances (e.g., different channel realizations, network sizes and QoS targets)
\begin{eqnarray}
\label{lss}
\mathscr{P}:
\mathop {\rm{minimize~}}_{{\bf{v}}\in\mathcal{V}} f({\bf{v}}),
\end{eqnarray}
where $f(\bf{v})$ is convex in $\bf{v}$ as shown in $\mathscr{P}_1({\bs{\gamma}})$ and $\mathscr{P}_2({\bs{\gamma}})$. Solving problem $\mathscr{P}$ means that the corresponding algorithm should return the optimal solution or the certificate of infeasibility.

For all the problems discussed above, problem $\mathscr{P}$ can be reformulated as an SOCP problem, and thus it can be solved in polynomial time via the interior-point method, which is implemented in most advanced off-the-shelf solvers, e.g., public software packages like SeDuMi \cite{SeDuMi_1999using} and SDPT3 \cite{SDPT3_1999} and commercial software packages like MOSEK \cite{Mosek_2000mosek}. However, the computational cost of such second-order methods will be prohibitive for large-scale problems. On the other hand, most custom algorithms, e.g., the uplink-downlink approach \cite{WeiYu_WC10} and the ADMM based algorithms \cite{shen2012distributed,Luo_2013base,joshi2014distributed}, however, fail to either scale well to large problem sizes or detect the infeasibility effectively.

To overcome the limitations of the scalability of the state-of-art solvers and the capability of infeasibility detection of the custom algorithms, in this paper, we propose to solve the homogeneous self-dual  embedding \cite{ye1994nl}
(which aims at providing necessary certificates) of problem $\mathscr{P}$
via a first-order optimization method \cite{boyd2011distributed} (i.e., the
operator splitting method). This will be presented in Section {\ref{opsp}}. To arrive at the homogeneous self-dual embedding and enable parallel computing, the original problem will be first transformed into a standard cone programming form as will be presented in Section {\ref{hsde}}. This forms the main idea of the two-stage based large-scale optimization framework as shown in Fig. {\ref{LSCO}}.

\section{Matrix Stuffing for Fast Standard Cone Programming Transformation}
\label{hsde}
Although the parser/solver modeling language framework, like CVX \cite{cvx}
and YALMIP \cite{YALMIP_2004yalmip}, can automatically transform the original
problem instance into a standard form, it requires substantial time to accomplish this procedure \cite{boyd_code2013, Yuanming_Globecom2014}. In particular, for each problem instance, the parser/solver modeling frameworks need to repeatedly parse and
canonicalize it. To avoid such modeling overhead of reading problem data
and repeatedly parsing and canonicalizing, we propose to use the matrix stuffing
technique \cite{boyd_code2013, Yuanming_Globecom2014} to perform fast transformation
by exploiting the problem structures.  Specifically, we will first
generate   the mapping from the original problem to the cone program, and then the structure of the standard form will be stored.
This can be accomplished offline. Therefore, for each problem instance, we only
need to stuff its parameters to data of the corresponding pre-stored structure of the standard cone program. Similar
ideas were presented in  the emerging parse/generator modeling frameworks like
CVXGEN \cite{CVXGEN_mattingley2012cvxgen} and QCML \cite{boyd_code2013}, which aim
at embedded applications for some specific problem families. In this paper,
we will  demonstrate in Section {\ref{num}} that matrix stuffing
is essential to scale to large problem sizes  for fast transformation at
the first stage of the proposed framework.

\subsection{Conic Formulation of Convex Programs}
In this section, we  describe a systematic way to transform the original problem
$\mathscr{P}$ to the standard cone program. To enable parallel computing, a common way is to replicate some variables through either exploiting problem structures \cite{shen2012distributed, Luo_2013base} or using the consensus formulation \cite{boyd2011distributed,joshi2014distributed}. However, directly working on these reformulations is difficult to provide computable mathematical certificates of infeasibility. Therefore, heuristic criteria are often adopted to detect the infeasibility, e.g., the underlying problem instance is reported
to be infeasible when the algorithm exceeds the pre-defined maximum iterations without convergence \cite{Luo_2013base}. To unify the requirements of parallel and scalable computing and to provide computable mathematical certificates of infeasibility, in this paper, we propose  to transform the original problem $\mathscr{P}$ to the following equivalent cone program $\mathscr{P}_{\textrm{cone}}$: \begin{eqnarray}
\label{soc}
\mathscr{P}_{\textrm{cone}}:
\mathop {\rm{minimize}}_{{\bs{\nu}}, {\bs{\mu}}}&& {\bf{c}}^{T}{\bs{\nu}}\nonumber\\
\label{affset}
{\rm{subject~to~}}&& {\bf{A}}{\bs{\nu}}+{\bs{\mu}}={\bf{b}}\\
\label{conset}
&& ({\bs{\nu}}, {\bs{\mu}})\in\mathbb{R}^{n}\times\mathcal{K},
\end{eqnarray}
where ${\bs{\nu}}\in\mathbb{R}^n$ and ${\bs{\mu}}\in\mathbb{R}^{m}$ are the
optimization variables, $\mathcal{K}=\{0\}^{r}\times\mathcal{S}^{m_1}\times\cdots\times\mathcal{S}^{m_q}$
with  $\mathcal{S}^{p}$ as
the standard second-order cone of dimension $p$
\begin{eqnarray}
\mathcal{S}^{p}=\{(y, {\bf{x}})\in\mathbb{R}\times\mathbb{R}^{p-1}|\|{\bf{x}}\|\le
y\},
\end{eqnarray}
and $\mathcal{S}^{1}$ is defined as the cone of nonnegative reals, i.e., $\mathbb{R}_{+}$. Here, each $\mathcal{S}^i$
has dimension $m_i$ such that $(r+\sum_{i=1}^q m_i)=m$, ${\bf{A}}\in\mathbb{R}^{m\times
n}$, ${\bf{b}}\in\mathbb{R}^{m}$, ${\bf{c}}\in\mathbb{R}^{n}$. The equivalence means that the optimal solution or the certificate of infeasibility of the original problem $\mathscr{P}$ can be extracted from the solution to the equivalent cone program $\mathscr{P}_{\textrm{cone}}$.  To reduce the storage and memory overhead, we store the matrix ${\bf{A}}$, vectors ${\bf{b}}$ and ${\bf{c}}$ in the sparse form \cite{Davis_sparselinear} by only storing the  non-zero entries. 

The general idea of such transformation is  to rewrite the original problem $\mathscr{P}$ into a Smith form by introducing a new variable for each subexpression in disciplined convex programming form \cite{boyd2008graph} of problem $\mathscr{P}$. The details are presented in the Appendix. Working with this transformed standard cone program $\mathscr{P}_{\textrm{cone}}$ has the following two advantages:
\begin{itemize}
\item The homogeneous self-dual embedding of the primal-dual pair of the standard cone program can be induced, thereby providing certificates of infeasibility. This will be presented in Section {\ref{hsd}}.
\item The feasible set $\mathcal{V}$ (\ref{intset}) formed by the intersection of a finite number of constraint sets ${\mathcal{P}}_l$'s and $\mathcal{Q}_k$'s in the original
problem $\mathscr{P}$ can be transformed into two sets in $\mathscr{P}_{\textrm{cone}}$: a subspace (\ref{affset}) and a convex cone $\mathcal{K}$, which is
formed by the Cartesian product of  second-order cones. This  salient
feature will be exploited to enable parallel and scalable computing, as will be presented
in Section {\ref{opsp1}}.
\end{itemize}

\subsection{Matrix Stuffing for Fast Transformation} 
Inspired by the work \cite{boyd_code2013} on fast optimization code deployment for embedding second-order cone program, we propose to use the matrix stuffing technique \cite{boyd_code2013, Yuanming_Globecom2014} to transform the original problem into the standard cone program quickly. Specifically, for any given network size, we first generate and store the structure that maps the original problem $\mathscr{P}$ to the standard form $\mathscr{P}_{\textrm{cone}}$. Thus, the pre-stored standard form structure includes the problem dimensions (i.e., $m$ and $n$), the description of $\mathcal{V}$ (i.e., the array of the cone sizes $[r, m_1,m_2,\dots, m_q]$), and the symbolic problem parameters ${\bf{A}}$, ${\bf{b}}$ and $\bf{c}$. This procedure can be done offline. 

Based on the pre-stored structure, for a given problem instance $\mathscr{P}$, we only need to copy  its parameters  (i.e., the channel coefficients ${\bf{h}}_K$'s, maximum transmit powers $P_{l}$'s, SINR
targets $\gamma_k$'s) to the corresponding data in the standard form $\mathscr{P}_{\textrm{cone}}$ (i.e., ${\bf{A}}$ and ${\bf{b}}$). Details of the exact description of copying data for transformation are presented in  the Appendix.
As the procedure for transformation only needs to copy memory, it thus is  suitable for fast transformation and can avoid repeated parsing and generating as in parser/solver modeling frameworks like CVX. 

\begin{remark}
As shown in the Appendix, the dimension of the transformed standard cone program $\mathscr{P}_{\textrm{cone}}$ becomes $m=(L+K)+(2NK+1)+\sum_{l=1}^{L}(2KN_l+1)\!+\!K(2K+2)$, 
which is much larger than the dimension of the original problem, i.e., $2NK$ in the equivalent real-field. But as discussed above, there are unique advantages of working with this standard form, which compensate for the increase in the size, as will be explicitly presented in later sections.      
\end{remark}

\section{The Operator Splitting Method For Large-Scale Homogeneous Self-Dual
Embedding}
\label{opsp}
Although the standard cone program $\mathscr{P}_{\textrm{cone}}$ itself is suitable for parallel computing via the operator splitting method \cite{Boyd_PD13}, directly working on this problem may fail to provide certificates of infeasibility. To address this limitation, based on the recent work by O'Donoghue {\emph{et al.}} \cite{Boyd_arXiv2013}, we propose to solve the homogeneous self-dual embedding \cite{ye1994nl} of the primal-dual pair of the  cone program $\mathscr{P}_{\textrm{cone}}$. The resultant homogeneous self-dual embedding is further solved via the operator splitting method, a.k.a. the ADMM algorithm \cite{boyd2011distributed}.

\subsection{Homogeneous Self-Dual Embedding of Cone Programming}
\label{hsd}
The basic idea of the homogeneous self-dual embedding is to embed the primal and dual problems of the cone program $\mathscr{P}_{\textrm{cone}}$ into a single feasibility problem (i.e., finding a feasible point of the intersection of a subspace and a convex set) such that either the  optimal solution or the certificate of infeasibility of the original cone program  $\mathscr{P}_{\textrm{cone}}$ can be extracted from the solution of the embedded problem.       

The dual problem of  $\mathscr{P}_{\textrm{cone}}$ is given by \cite{Boyd_arXiv2013}
\begin{eqnarray}
\label{dcone}
\mathscr{D}_{\textrm{cone}}:
\mathop {\rm{maximize}}_{{\boldsymbol{\eta}}, {\boldsymbol{\lambda}}}&&-{\bf{b}}^{T}{\boldsymbol{\eta}}\nonumber\\
{\rm{subject~to}}&& -{\bf{A}}^{T}{\boldsymbol{\eta}}+{\boldsymbol{\lambda}}={\bf{c}}\nonumber\\
&& ({\boldsymbol{\lambda}}, {\boldsymbol{\eta}})\in\{0\}^{n}\times\mathcal{K}^{*},
\end{eqnarray}
where $\boldsymbol{\lambda}\in\mathbb{R}^{n}$ and $\boldsymbol{\eta}\in\mathbb{R}^m$
are the dual variables, $\mathcal{K}^*$ is the dual cone of the convex cone
$\mathcal{K}$. Note that $\mathcal{K}=\mathcal{K}^{*}$, i.e., $\mathcal{K}$ is self dual.
Define the optimal values of the primal program $\mathscr{P}_{\textrm{cone}}$
and dual program $\mathscr{D}_{\textrm{cone}}$ are $p^{\star}$ and $d^{\star}$,
respectively.  Let $p^{\star}=+\infty$
and $p^{\star}=-\infty$ indicate primal infeasibility and unboundedness,
respectively. Similarly, let $d^{\star}=-\infty$ and $d^{\star}=+\infty$
indicate the dual infeasibility and unboundedness, respectively. We assume  strong duality for the convex cone program $\mathscr{P}_{\textrm{cone}}$, i.e., $p^{\star}=d^{\star}$, including  cases when they are infinite. This is a standard assumption for practically designing solvers for conic programs, e.g., it is assumed in \cite{SeDuMi_1999using,SDPT3_1999,Mosek_2000mosek,ye1994nl,Boyd_arXiv2013}. Besides, we do not make any regularity assumption on
the feasibility and boundedness assumptions on the primal and dual problems.   

\subsubsection{Certificates of Infeasibility}
Given the cone program $\mathscr{P}_{\textrm{cone}}$, a main task is to detect feasibility. In \cite[Theorem
1]{Shamai_SP2006}, a {sufficient} condition for the existence
of strict feasible solution was provided for the transmit power minimization problem without power constraints. However, for the general problem $\mathscr{P}$ with per-MU QoS constraints and per-RAU transmit power constraints, it is difficult to obtain such a feasibility condition analytically. Therefore, most existing works either assume that the underlying problem is feasible \cite{WeiYu_WC10} or provide  heuristic ways to handle infeasibility \cite{Luo_2013base}.    

Nevertheless, the only way to detect infeasibility effectively is to provide a certificate or proof of infeasibility as presented in the following proposition.  \begin{proposition}
\label{prop1}
[Certificates of Infeasibility] The following system
\begin{eqnarray}
{\bf{A}}{\bs{\nu}}+{\bs{\mu}}={\bf{b}}, {\bs{\mu}}\in\mathcal{K},
\end{eqnarray} 
is infeasible if and only if the following system is feasible
\begin{eqnarray}
\label{cid}
{\bf{A}}^{T}{\boldsymbol{\eta}}={\bf{0}}, {\bs{\eta}}\in\mathcal{K}^{\star}, {\bf{b}}^{T}{\boldsymbol{\eta}}<0.
\end{eqnarray}
Therefore, any dual variable ${\bs{\eta}}$ satisfying the system (\ref{cid}) provides a certificate or proof that the primal program $\mathscr{P}_{\textrm{cone}}$ (equivalently the original problem $\mathscr{P}$) is infeasible.

Similarly, any primal variable ${\bs{\nu}}$ satisfying the following system
\begin{eqnarray}
-{\bf{A}}{\bs{\nu}}\in\mathcal{K}, {\bf{c}}^{T}{\bs{\nu}}<0,
\end{eqnarray}
is a certificate of the dual program $\mathscr{D}_{\textrm{cone}}$ infeasibility. 
\begin{IEEEproof}
This result directly follows the theorem of strong alternatives \cite[Section 5.8.2]{boyd2004convex}.
\end{IEEEproof}
\end{proposition}
  
\subsubsection{Optimality Conditions}
If the transformed standard cone program $\mathscr{P}_{\textrm{cone}}$ is feasible, then (${\bs{\nu}}^{\star}, {\bs{\mu}}^{\star}, {\bs{\lambda}}^{\star}, {\bs{\eta}}^{\star}$) are optimal if and only if they satisfy the following Karush-Kuhn-Tucker (KKT) conditions
\begin{eqnarray}
\label{kkt1}
{\bf{A}}{\bs{\nu}}^{\star}+{\bs{\mu}}^{\star}-{\bf{b}}={\bf{0}}\\
\label{kkt2}
{\bf{A}}^{T}{\boldsymbol{\eta}}^{\star}- {\boldsymbol{\lambda}}^{\star}+{\bf{c}}={\bf{0}}\\
\label{kkt3}
({\boldsymbol{\eta}}^{\star})^{T}{\bs{\mu}}^{\star}=0\\
\label{kkt4}
({\bs{\nu}}^{\star}, {\bs{\mu}}^{\star}, {\boldsymbol{\lambda}}^{\star},
{\boldsymbol{\eta}}^{\star})\in\mathbb{R}^{n}\times\mathcal{K}\times\{0\}^{n}\times\mathcal{K}^{*}.
\end{eqnarray}
In particular, the complementary slackness condition (\ref{kkt3}) can be rewritten as
\begin{eqnarray}
{\bf{c}}^{T}{\bs{\nu}}^{\star}+{\bf{b}}^{T}{\boldsymbol{\eta}}^{\star}=0,
\end{eqnarray}
which explicitly forces the duality gap to be zero.

\subsubsection{Homogeneous Self-Dual Embedding}
We can first detect feasibility by Proposition {\ref{prop1}}, and then solve the KKT system if the problem is feasible and bounded. However, the disadvantage of such a two-phase method is that two related problems (i.e., checking feasibility and solving KKT conditions) need to be solved sequentially \cite{ye1994nl}. To avoid such inefficiency, we propose to solve the following homogeneous self-dual embedding
\cite{ye1994nl}:
\begin{eqnarray}
{\bf{A}}{\bs{\nu}}+{\bs{\mu}}-{\bf{b}}\tau={\bf{0}}\\
{\bf{A}}^{T}{\boldsymbol{\eta}}- {\boldsymbol{\lambda}}+{\bf{c}}\tau={\bf{0}}\\
{\bf{c}}^{T}{\bs{\nu}}+{\bf{b}}^{T}{\boldsymbol{\eta}}+\kappa={\bf{0}}\\
\label{hsd4}
\!\!\!\!\!\!\!\!({\bs{\nu}}, {\bs{\mu}}, {\boldsymbol{\lambda}},
{\boldsymbol{\eta}}, \tau, \kappa)\in
\mathbb{R}^{n}\times\mathcal{K}\times\{0\}^{n}\times\mathcal{K}^{*}\times\mathbb{R}_{+}\times\mathbb{R}_{+},
\end{eqnarray}
to embed all the information on the infeasibility and optimality into a single
system by introducing two new nonnegative variables ${{\tau}}$ and ${{\kappa}}$, which encode different outcomes. The homogeneous self-dual embedding thus can be rewritten as the following compact form 
\begin{eqnarray}
\label{hsdf}
\mathscr{F}_{\textrm{HSD}}:
\mathop {\rm{find}}&&({\bf{x}}, {\bf{y}})\nonumber\\
{\rm{subject~to}}&& {\bf{y}}={\bf{Q}}{\bf{x}}\nonumber\\
&& {\bf{x}}\in\mathcal{C}, {\bf{y}}\in\mathcal{C}^{*},
\end{eqnarray}
where 
\begin{eqnarray}
\underbrace{\left[ \begin{array}{c}
{\bs{\lambda}} \\
{\boldsymbol{\mu}} \\
{{\kappa}}
\end{array} \right]}_{\bf{y}}=\underbrace{\left[ \begin{array}{ccc}
{\bf{0}} & {\bf{A}}^{T} & {\bf{c}} \\
-{\bf{A}} & {\bf{0}} & {\bf{b}} \\
-{\bf{c}}^{T} & -{\bf{b}}^{T} & {\bf{0}}
\end{array} \right]}_{\bf{Q}}
\underbrace{\left[ \begin{array}{c}
{\bs{\nu}} \\
{\boldsymbol{\eta}} \\
{{\tau}}
\end{array} \right]}_{\bf{x}}, 
\end{eqnarray}
${\bf{x}}\in\mathbb{R}^{m+n+1}$, ${\bf{y}}\in\mathbb{R}^{m+n+1}$, ${\bf{Q}}\in\mathbb{R}^{(m+n+1)\times
(m+n+1)}$, $\mathcal{C}=\mathbb{R}^n\times
\mathcal{K}^*\times 
\mathbb{R}_+$ and $\mathcal{C}^{*}=\{0\}^n\times\mathcal{K}\times\mathbb{R}_{+}$. This system has a trivial solution with all
variables as zeros.

The homogeneous self-dual embedding problem $\mathscr{F}_{\textrm{HSD}}$
is thus a feasibility problem finding a nonzero solution in the intersection of a subspace and a convex cone.  
Let $({\bs{\nu}}, {\bs{\mu}}, {\boldsymbol{\lambda}}, {\boldsymbol{\eta}}, \tau, \kappa)$ be a non-zero solution of the  homogeneous self-dual embedding. We then have the following remarkable trichotomy derived in \cite{ye1994nl}: 
\begin{itemize}
\item {\textbf{Case 1}}: $\tau>0$, $\kappa=0$, then
\begin{eqnarray}
\hat{\bs{\nu}}={\bs{\nu}}/\tau, \hat{\bs{\eta}}={\bs{\eta}}/\tau, \hat{\bs{\mu}}={\bs{\mu}}/\tau
\end{eqnarray}
are the primal and dual solutions to the cone program $\mathscr{P}_{\textrm{cone}}$.

\item {\textbf{Case 2}}:  $\tau=0$, $\kappa>0$; this implies ${\bf{c}}^{T}{\bs{\nu}}+{\bf{b}}^{T}{\boldsymbol{\eta}}<0$, then
\begin{enumerate}
\item If ${\bf{b}}^{T}{\boldsymbol{\eta}}<0$, then $\hat{\bs{\eta}}={\bs{\eta}}/(-{\bf{b}}^{T}{\boldsymbol{\eta}})$ is a  certificate of the primal infeasibility as
\begin{eqnarray}
{\bf{A}}^{T}\hat{\boldsymbol{\eta}}={\bf{0}}, \hat{\bs{\eta}}\in\mathcal{V}^{\star},
{\bf{b}}^{T}\hat{\boldsymbol{\eta}}=-1.
\end{eqnarray}
\item If ${\bf{c}}^{T}{\bs{\nu}}<0$, then $\hat{\bs{\nu}}={\bs{\nu}}/(-{\bf{c}}^{T}\hat{\bs{\nu}})$ is a certificate of the dual infeasibility as
\begin{eqnarray}
-{\bf{A}}\hat{\bs{\nu}}\in\mathcal{V}, {\bf{c}}^{T}\hat{\bs{\nu}}=-1. 
\end{eqnarray}
\end{enumerate}
\item {\textbf{Case 3}}: $\tau=\kappa=0$; no conclusion can be made about the cone problem $\mathscr{P}_{\textrm{cone}}$. 
\end{itemize}

Therefore, from the solution to the homogeneous self-dual embedding, we can extract either the optimal solution (based on (\ref{solv})) or the certificate of infeasibility for the original problem. Furthermore, as the set (\ref{hsd4}) is a Cartesian product of a finite number of sets, this will enable parallelizable algorithm design. With the distinct advantages of the homogeneous self-dual embedding, in the sequel, we focus on developing efficient algorithms to solve the large-scale feasibility problem $\mathscr{F}_{\textrm{HSD}}$ via the operator splitting method.

\subsection{The Operator Splitting Method}
\label{opsp1}
Conventionally, the convex homogeneous self-dual embedding $\mathscr{F}_{\textrm{HSD}}$ can be solved via the interior-point method, e.g., \cite{SeDuMi_1999using,SDPT3_1999,Mosek_2000mosek,ye1994nl}. However, such second-order method has cubic computational complexity for the second-order cone programs \cite{nesterov1994interior}, and thus the computational cost will be prohibitive for large-scale problems. Instead, O'Donoghue {\emph{et al.}} \cite{Boyd_arXiv2013} develop a first-order optimization algorithm based on   the operator splitting method, i.e., the ADMM algorithm \cite{boyd2011distributed}, to solve the large-scale homogeneous self-dual embedding. The key observation is that the convex cone constraint in $\mathscr{F}_{\textrm{HSD}}$ is the Cartesian product of standard convex cones (i.e., second-order cones, nonnegative reals and free variables), which enables parallelizable computing. Furthermore, we will show that the computation of each iteration in the operator splitting method is very cheap and efficient. 

Specifically, the homogeneous self-dual  embedding  $\mathscr{F}_{\textrm{HSD}}$ can be rewritten as
\begin{eqnarray}
\label{hsdf1}
\mathop {\rm{minimize}}&&I_{\mathcal{C}\times\mathcal{C}^{*}}({\bf{x}}, {\bf{y}})+I_{{\bf{Q}}{\bf{x}}={\bf{y}}}({\bf{x}}, {\bf{y}}),
\end{eqnarray} 
where $I_{\mathcal{S}}$ is the indicator function of the set $\mathcal{S}$, i.e., $I_{\mathcal{S}}(z)$ is zero for $z\in\mathcal{S}$ and it is $+\infty$ otherwise.
By replicating  variables ${\bf{x}}$ and ${\bf{y}}$, problem (\ref{hsdf1}) can be transformed into the following consensus form \cite[Section 7.1]{boyd2011distributed}
\begin{eqnarray}
\label{admm}
\mathscr{P}_{\textrm{ADMM}}: 
\mathop {\rm{minimize}}&&
I_{\mathcal{C}\times\mathcal{C}^{*}}({\bf{x}}, {\bf{y}})+I_{{\bf{Q}}\tilde{\bf{x}}=\tilde{\bf{y}}}(\tilde{\bf{x}},
\tilde{\bf{y}})\nonumber\\
{\rm{subject~to}}&& ({\bf{x}}, {\bf{y}})=(\tilde{\bf{x}}, \tilde{\bf{y}}),
\end{eqnarray} 
which is readily to be solved by the operator splitting method. 

Applying the ADMM algorithm \cite[Section 3.1]{boyd2011distributed} to  problem $\mathscr{P}_{\textrm{ADMM}}$ and eliminating the dual variables  by exploiting the self-dual property of the problem $\mathscr{F}_{\textrm{HSD}}$ (Please refer to \cite[Section 3]{Boyd_arXiv2013} on how to simplify the ADMM algorithm),
the final algorithm is shown as follows:
\begin{eqnarray}
\label{proxop}
{\mathcal{OS}}_{\textrm{ADMM}}:\left\{\begin{array}{l}
\tilde{\bf{x}}^{[i+1]}=({\bf{I}}+{\bf{Q}})^{-1}({\bf{x}}^{[i]}+{\bf{y}}^{[i]})\\
{\bf{x}}^{[i+1]}=\Pi_{\mathcal{C}}(\tilde{\bf{x}}^{[i+1]}-{\bf{y}}^{[i]})\\
{\bf{y}}^{[i+1]}={\bf{y}}^{[i]}-\tilde{\bf{x}}^{[i+1]}+{\bf{x}}^{[i+1]},
\end{array}\right.
\end{eqnarray} 
where $\Pi_{\mathcal{C}}({\bf{x}})$ denotes the Euclidean projection of $\bf{x}$
onto the set $\mathcal{C}$. This algorithm has the  $\mathcal{O}(1/k)$ convergence rate \cite{Brendan_SIAM2014fastADMM} with $k$ as the iteration counter (i.e., the $\epsilon$ accuracy can be achieved in $\mathcal{O}(1/\epsilon)$ iterations) and will not converge to zero if a nonzero solution exists \cite[Section 3.4]{Boyd_arXiv2013}. Empirically, this algorithm can converge to modest accuracy within a reasonable amount of time. As the last step is computationally trivial,  in the sequel, we will focus on how to solve the first two steps efficiently.

\subsubsection{Subspace Projection via Factorization Caching}
The first step in the algorithm ${\mathcal{OS}}_{\textrm{ADMM}}$ is a subspace projection. After simplification \cite[Section 4]{Boyd_arXiv2013}, we essentially need to solve the following linear equation at each iteration, i.e., \begin{eqnarray}
\label{lin}
\underbrace{\left[ \begin{array}{cc}
{\bf{I}} & -{\bf{A}}^{T} \\
-{\bf{A}} & -{\bf{I}}
\end{array} \right]}_{\bf{S}}\underbrace{\left[ \begin{array}{c}
{\bs{\nu}} \\
{-\boldsymbol{\eta}}
\end{array} \right]}_{\bf{x}}=\underbrace{\left[ \begin{array}{c}
{\bs{\nu}}^{[i]} \\
{\boldsymbol{\eta}}^{[i]} 
\end{array} \right]}_{\bf{b}},
\end{eqnarray}  
for the given ${\bs{\nu}}^{[i]}$ and ${\bs{\eta}}^{[i]}$ at iteration $i$, where ${\bf{S}}\in\mathbb{R}^{d\times d}$ with $d=m+n$ is a \emph{symmetric quasidefinite} matrix \cite{Vanderbei}. To enable quicker inversions and reduce memory overhead via exploiting the sparsity of the matrix $\bf{S}$, the sparse permuted ${\sf{LDL}}^{T}$ factorization \cite{Davis_sparselinear} method can be adopted. Specifically, such factor-solve method can be carried out by first computing the sparse permuted ${\sf{LDL}}^{T}$ factorization as follows
\begin{eqnarray}
\label{factor}
{\bf{S}}={\bf{PLDL}}^{T}{\bf{P}}^{T},
\end{eqnarray}  
where $\bf{L}$ is a lower triangular matrix, $\bf{D}$ is a diagonal matrix \cite{Boyd_PD13} and  $\bf{P}$ with ${\bf{P}}^{-1}={\bf{P}}^{T}$ is a permutation matrix to fill-in of the factorization \cite{Davis_sparselinear}, i.e., the number of nonzero entries in ${\bf{L}}$. Such factorization exists for any permutation ${\bf{P}}$, as the matrix $\bf{S}$ is symmetric quasidefinite \cite[Theorem 2.1]{Vanderbei}. Computing the factorization costs much less than $\mathcal{O}(1/3 d^3)$ flops, while the exact value depends on $d$ and the sparsity pattern of ${\bf{S}}$ in a complicated way. Note that such factorization only needs to be computed once in the first iteration and can be cached for re-using in the sequent iterations for subspace projections. This is called the \emph{factorization caching} technique \cite{Boyd_arXiv2013}. 

Given the cached factorization (\ref{factor}), solving subsequent projections ${\bf{x}}={\bf{S}}^{-1}{\bf{b}}$ (\ref{lin}) can be carried out by solving the following much easier equations:
\begin{eqnarray}
\!\!\!\!\!\!\!{\bf{P}}{\bf{x}}_1={\bf{b}}, {\bf{L}}{\bf{x}}_2={\bf{x}}_1, {\bf{D}}{\bf{x}}_3={\bf{x}}_2, {\bf{L}}^{T}{\bf{x}}_4={\bf{x}}_3, {\bf{P}}^{T}{\bf{x}}={\bf{x}}_{4},
\end{eqnarray}
which cost zero flops, $\mathcal{O}(sd)$ flops by forward substitution with $s$ as the number of nonzero entries in $\bf{L}$, $\mathcal{O}(d)$ flops, $\mathcal{O}(sd)$ flops by backward substitution, and zero flops, respectively \cite[Appendix C]{boyd2004convex}.   

\subsubsection{Cone Projection via Proximal Operator Evaluation}
The second step in the algorithm $\mathcal{OS}_{\textrm{ADMM}}$ is to project a point ${\bs{\omega}}$
onto the cone $\mathcal{C}$. As $\mathcal{C}$ is the Cartesian product of
the finite number of convex cones $\mathcal{C}_i$, we can perform projection onto $\mathcal{C}$ by projecting
onto $\mathcal{C}_i$ separately and in parallel. Furthermore, the projection
onto  each convex cone can be done with closed-forms. Specifically, for nonnegative real $\mathcal{C}_i=\mathbb{R}_{+}$, 
we have that \cite[Section
6.3.1]{Boyd_2013proximal}
\begin{eqnarray}
\label{pronon}
\Pi_{\mathcal{C}_i}({\boldsymbol{\omega}})={\boldsymbol{\omega}}_+,
\end{eqnarray}
where the nonnegative part operator $(\cdot)_+$ is taken elementwise. For
the second-order cone ${\mathcal{C}}_i=\{(y, {\bf{x}})\in\mathbb{R}\times\mathbb{R}^{p-1}|
\|{\bf{x}}\|\le y\}$, we have that \cite[Section 6.3.2]{Boyd_2013proximal}
\begin{eqnarray}
\label{proxop}
\Pi_{\mathcal{C}_i}({\boldsymbol{\omega}},\tau)=\left\{\begin{array}{l}
0, \|{\boldsymbol{\omega}}\|_2\le -\tau\\
({\boldsymbol{\omega}},\tau), \|{\boldsymbol{\omega}}\|_2\le \tau\\
(1/2)(1+\tau/\|{\boldsymbol{\omega}}\|_2)({\boldsymbol{\omega}},\|{\boldsymbol{\omega}}\|_2),
\|{\boldsymbol{\omega}}\|_2\ge |\tau|.
\end{array}\right.
\end{eqnarray} 

In summary, we have presented that each step in the algorithm ${\mathcal{OS}}_{\textrm{ADMM}}$ can be computed efficiently. In particular, from both (\ref{pronon}) and (\ref{proxop}), we see that the cone projection
can be carried out very efficiently with closed-forms, leading to parallelizable algorithms.

\section{Practical Implementation Issues}
\label{pii}
In previous sections, we have presented the unified two-stage framework for large-scale convex optimization in dense wireless cooperative networks. In this section, we will focus on the implementation issues of the proposed framework. 

\subsection{Automatic Code Generation for Fast Transformation}
In the Appendix, we describe a systematic way to transform the original problem to the standard cone programming form. The resultant structure that maps the original problem to the standard form can be stored and re-used for fast transforming via matrix stuffing. This can significantly reduce the modeling overhead compared with the parse/solver modeling frameworks like CVX. However, it requires tedious manual works to find the mapping and may not be easy to verify the correctness of the generated mapping.  Chu \emph{et al.} \cite{boyd_code2013} gave such an attempt intending to automatically generate the code for matrix stuffing. However, the corresponding software package QCML \cite{boyd_code2013}, so far, is far from complete and may not be suitable for our applications. Extending
the numerical-based transformation modeling  frameworks like CVX to the symbolic-based transformation
modeling frameworks like QCML is not trivial and requires tremendous mathematical and technical efforts. In this paper, we derive the mapping in the Appendix manually and verify the correctness by comparing with CVX through extensive simulations.            

\subsection{Implementation of the Operator Splitting Algorithm}
\label{cfos}
Theoretically,  the presented operator splitting algorithm $\mathcal{OS}_{\textrm{ADMM}}$ is compact, {parameter-free}, with parallelizable computing and linear convergence. Practically, there are typically several ways to improve the efficiency of the algorithm. In particular, there are various tricks that can be employed to improve the convergence rate, e.g, over-relaxation, warm-staring and problem data scaling as described in \cite{Boyd_arXiv2013}. In the dense wireless cooperative networks with multi-entity collaborative architecture, we are  interested in two particular ways to speed up the subspace projection of the algorithm $\mathcal{OS}_{\textrm{ADMM}}$, which is the main computational bottleneck. Specifically, one way is to use the parallel algorithms for the factorization (\ref{factor}) by utilizing the distributed computing and memory resources \cite{Poulson2013elemental}. For instance, in the cloud computing environments in Cloud-RAN, all the baseband units share the computing, memory and storage resources in a single baseband unit pool \cite{mobile2011c, Yuanming_TWC2014}. Another way is to leverage \emph{symbolic} factorization (\ref{factor}) to speed up the numerical factorization for each problem instance, which is a general idea for the code generation system CVXGEN \cite{CVXGEN_mattingley2012cvxgen} for realtime convex quadratic optimization \cite{Boyd_2010real} and the interior-point method based SOCP solver \cite{Boyd_2013ecos} for embedded systems. Eventually, the ADMM solver in Fig. {\ref{LSCO}} can be symbolic based so as to provide numerical solutions for each problem instance extremely fast and in a realtime way. However, this requires further investigation.

\section{Numerical Results}
\label{num}
In this section, we simulate the proposed two-stage  based large-scale convex optimization framework for  performance optimization in dense wireless cooperative networks. \emph{The corresponding MATLAB code that can reproduce all the simulation results using the proposed large-scale convex optimization algorithm is available online\footnote{https://github.com/SHIYUANMING/large-scale-convex-optimization}}. 

We consider the following channel model for the link between the $k$-th MU and the
$l$-th RAU:
\begin{eqnarray}
{\bf{h}}_{kl}=10^{-L(d_{kl})/20}\sqrt{\varphi_{kl}s_{kl}}{\bf{f}}_{kl},
\forall k,l,
\end{eqnarray}
where $L(d_{kl})$ is the path-loss in dB at distance $d_{kl}$ as shown in \cite[Table I]{Yuanming_TWC2014}, $s_{kl}$ is the shadowing coefficient, $\varphi_{kl}$
is the antenna gain and ${\bf{f}}_{kl}$ is the small-scale fading coefficient.
We use the standard cellular network parameters as showed in \cite[Table
I]{Yuanming_TWC2014}. All the simulations are carried out on a personal computer
with 3.2 GHz quad-core Intel Core i5 processor and 8 GB of RAM running  Linux. The reference implementation of the operator splitting algorithm SCS is  available online\footnote{https://github.com/cvxgrp/scs}, which is a general software package for solving large-scale convex cone problems based on \cite{Boyd_arXiv2013} and can be called by the modeling frameworks CVX and CVXPY \cite{cvxpy}. The settings (e.g., the stopping criteria) of SCS can be found in \cite{Boyd_arXiv2013}. 

The proposed two-stage approach framework, termed ``{\sf{Matrix~Stuffing+SCS}}", is compared with the following state-of-art frameworks:
\begin{itemize}
\item {\sf{CVX+SeDuMi/SDPT3/MOSEK}}: This category adopts second-order methods.
The modeling framework CVX  will first automatically transform the original problem
instance (e.g., the problem $\mathscr{P}$ written in the disciplined convex
programming form) into the standard cone programming form and then call
an interior-point  solver, e.g., SeDuMi \cite{SeDuMi_1999using}, SDPT3 \cite{SDPT3_1999} or MOSEK \cite{Mosek_2000mosek}.
 
\item {\sf{CVX+SCS}}: In this first-order method based framework, CVX first transforms
the original problem instance into the standard form and then calls the operator
splitting solver SCS. 
\end{itemize}

We define the ``{\bfseries{modeling time}}" as the transformation time for the first stage, the ``{\bfseries{solving time}}" as the time spent for the second stage, and the ``{\bfseries{total time}}" as the time of the two stages for solving one problem instance. As the large-scale convex optimization algorithm should scale well to both the modeling part and the solving part simultaneously, the time comparison of each individual stage will demonstrate the effectiveness of the proposed two-stage approach.  

Given the network size,  we first generate and store the problem
structure of the standard form $\mathscr{P}_{\textrm{cone}}$, i.e., the
structure of ${\bf{A}}$, ${\bf{b}}$, $\bf{c}$ and the descriptions of $\mathcal{V}$.
As this procedure can be done offline for all the candidate network sizes, we thus ignore this step for time comparison.
We repeat the following procedures to solve the large-scale convex optimization
problem $\mathscr{P}$ with different parameters and sizes using the proposed framework ``{\sf{Matrix Stuffing+SCS}}":
\begin{enumerate}
\item Copy the parameters in the problem instance $\mathscr{P}$ to the data  in the pre-stored structure of the standard cone program $\mathscr{P}_{\textrm{cone}}$. 
\item Solve the resultant standard cone programming instance  $\mathscr{P}_{\textrm{cone}}$ using the solver {{SCS}}.
\item Extract the optimal solutions of $\mathscr{P}$ from the solutions to  $\mathscr{P}_{\textrm{cone}}$ by the solver SCS. 
\end{enumerate}      

Finally, note that all the interior-point solvers are multiple threaded (i.e., they can utilize multiple threads to gain extra speedups), while the operator splitting algorithm solver SCS is single threaded. Nevertheless, we will show that SCS performs much faster than the interior-point solvers. We also emphasize that the operator splitting method aims at  scaling well to large problem sizes and thus provides solutions
to modest accuracy within reasonable time, while the interior-point
method intends to provide highly accurate solutions. Furthermore, the modeling
framework CVX aims at rapid prototyping and providing a user-friendly tool
for automatically transformations for  general problems, while the matrix-stuffing technique targets at scaling to large-scale problems for the specific problem family $\mathscr{P}$. Therefore, these frameworks and solvers are not really comparable with different purposes and application
capabilities. We mainly use them to verify the effectiveness and reliability of our proposed
framework in terms of the solution time and the solution quality.

\begin{table*}[!t]
\renewcommand{\arraystretch}{1.3}
\caption{Time and Solution Results for Different Convex Optimization Frameworks}
\label{mats_table}
\centering
\begin{tabular}{c|c|c|c|c|c|c}
{{Network Size ($L=K$)}} &  & 20 & 50 & 100 & 150 & 200 \\
\hline
\multirow{2}{*}{\sf{CVX+SeDuMi}}&  {\bfseries{Total Time}} [sec] & $\bf{8.1164}$ & N/A& N/A & N/A & N/A \\
\cline{2-7}
 &  Objective [W]& 12.2488 & N/A & N/A & N/A & N/A\\
\hline\hline
\multirow{2}{*}{\sf{CVX+SDPT3}}&  {\bfseries{Total Time}} [sec] & $\bf{5.0398}$ &  ${\bf{330.6814}}$ & N/A &
N/A & N/A \\
\cline{2-7}
 &  Objective [W]& 12.2488 & 6.5216 & N/A & N/A & N/A\\
\hline\hline
\multirow{2}{*}{\sf{CVX+MOSEK}}&  {\bfseries{Total Time}} [sec] & ${\bf{1.2072}}$  & ${\bf{51.6351}}$ & N/A &
N/A & N/A \\
\cline{2-7}
 &  Objective [W]& 12.2488 & 6.5216 & N/A & N/A & N/A \\
\hline\hline
\multirow{3}{*}{\sf{CVX+SCS}}&  {\bfseries{Total Time}} [sec] & $\bf{0.8501}$ & ${\bf{5.6432}}$ & ${\bf{51.0472}}$ & $\bf{227.9894}$
 & $\bf{725.6173}$ \\
\cline{2-7}
 &  {\bfseries{Modeling Time}} [sec]& 0.7563 & 4.4301 & 38.6921 & 178.6794 & 534.7723\\
 \cline{2-7}
  &  Objective [W]& 12.2505 & 6.5215 & 3.1303 & 2.0693 & 1.5404\\
\hline\hline
\multirow{3}{*}{\sf{Matrix Stuffing+SCS}}&  {\bfseries{Total Time}} [sec] & $\bf{0.1137}$ & ${\bf{2.7222}}$ & ${\bf{26.2242}}$ & ${\bf{90.4190}}$
 & $\bf{328.2037}$ \\
\cline{2-7}
 &  {\bfseries{Modeling Time}} [sec]& 0.0128 & 0.2401 & 2.4154 & 9.4167 & 29.5813\\
 \cline{2-7}
  &  Objective [W]& 12.2523 & 6.5193 & 3.1296 & 2.0689 & 1.5403\\
\hline
\end{tabular}
\label{timecom}
\vspace*{-10pt}
\end{table*} 

\subsection{Effectiveness and Reliability of the Proposed Large-Scale Convex Optimization
Framework}
Consider a network with $L $ $2$-antenna RAUs and
$K$ single-antenna MUs uniformly and independently
distributed{\footnote{\rev{Consider the CSI acquisition overhead, our proposed approach is mainly
suitable in the low user mobility scenarios}. }} in the square region $[-3000, 3000] \times [-3000, 3000]$
meters with $L=K$. We consider the total transmit power minimization problem $\mathscr{P}_1({\bs{\gamma}})$ with the  QoS requirements for each MU as $\gamma_k=5$ dB, $\forall k$. Table {\ref{timecom}} demonstrates the comparison of the running time and solutions using different convex optimization frameworks.  Each point of the simulation results is averaged over $100$ randomly generated network realizations (i.e., one small scaling fading realization for each large-scale fading realization).  

For the modeling time comparisons, this table shows that  the value of the proposed matrix stuffing technique ranges between 0.01 and 30 seconds{\footnote{This value can  be significantly reduced in practical implementations, e.g., at the BBU pool in Cloud-RAN, which, however, requires substantial further investigation. Meanwhile, the results effectively confirm that the proposed matrix stuffing technique scales well to large-scale problems. }} for different network sizes and can speedup about 15x to 60x compared to the parser/solver modeling framework CVX. In particular, for large-scale problems, the transformation using CVX is time consuming and becomes the bottleneck, as the ``{\bfseries{modeling time}}" is comparable and even larger than the ``{\bfseries{solving time}}". For example, when $L=150$, the ``{\bfseries{modeling time}}" using CVX is about 3 minutes, while the matrix stuffing only requires about 10 seconds. Therefore, the matrix stuffing for fast transformation is essential for solving large-scale convex optimization problems quickly. 

For the solving time (which can be easily calculated by subtracting the ``{\bfseries{modeling time}}" from the ``{\bfseries{total time}}") using different solvers, this table shows that the operator splitting solver can speedup by several orders of magnitude over the interior-point solvers. For example, for $L=50$, it can speedup about 20x and 130x over  MOSEK{\footnote{Although
SeDuMi, SDPT3 and MOSEK (commercial software) are all based on the interior-point
method, the implementation efficiency of the corresponding software packages
varies substantially. In the following simulations, we mainly compare with
the state-of-art public solver SDPT3.}} and SDPT3, respectively, while SeDuMi is inapplicable for this problem size as the running time exceeds the pre-defined maximum value, i.e., one hour. In particular, all the interior-point solvers fail to solve large-scale problems (i.e., $L=100, 150 ,200$), denoted as ``N/A", while the operator splitting solver SCS can scale well to large problem sizes. For the largest problems with $L=200$, the operator splitting  solver  can solve them in   about 5 minutes.

For the quality of the solutions, this table shows that the propose framework can provide a solution to modest accuracy within much less time. For the two problem sizes, i.e., $L=20$ and $L=50$, which can be solved by the interior-point method based frameworks, the optimal values attained by the proposed framework are within $0.03\%$ of that obtained via the second-order method frameworks.

In summary, the proposed two-stage  based large-scale convex optimization framework scales well to large-scale problem modeling and solving simultaneously. Therefore, it provides an effective way to evaluate the system performance  via large-scale optimization in dense wireless networks. However, its implementation and performance in practical systems still need further investigation.  In particular, this set of results indicate that the scale of cooperation in dense wireless networks may be fundamentally constrained by the computation
complexity/time. 
\subsection{Infeasibility Detection Capability}
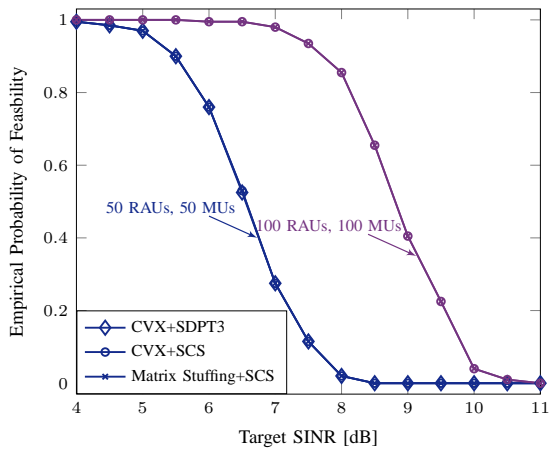
\begin{figure}[t]
\centering
\begin{tikzpicture}[scale=0.9]
\begin{axis}[
xmax=11, xmin=4, ymax=1.03, ymin=-0.03, 
xlabel={Target SINR [dB]}, 
ylabel={Empirical Probability of Feasbility}, 
every axis x label/.style={at={(ticklabel cs:0.5)},anchor=near
ticklabel},
every axis y label/.style={at={(ticklabel cs:0.5)},rotate=90, anchor=near
ticklabel}, 
label style={font=\footnotesize},
tick label style={font=\scriptsize},
legend style={at={(0,0)}, anchor=south west, font=\scriptsize}]
\addplot[colorhkust, mark=diamond, mark size=3pt, line width=0.8pt] table {feasibile_small_SDPT3.dat};
\addplot[colorhkust, mark=o, mark size=1.6pt, line width=0.8pt] table {feasibile_small_SCS.dat};
\addplot[colorhkust, mark=x, mark size=1.6pt, line width=0.8pt] table {feasibile_small_matscs.dat};

\addplot[colortsinghua, mark=o, mark size=1.6pt, line width=0.8pt] table {feasibile_large_SCS.dat};
\addplot[colortsinghua, mark=x, mark size=1.6pt, line width=0.8pt] table {feasibile_large_matscs.dat};

\draw [-latex', colorhkust, line width=0.5pt] (axis cs:6,0.46) --  (axis cs:6.73,0.4) node[above left=2mm] {\scriptsize{50 RAUs, 50 MUs}}; 
\draw [-latex', colortsinghua, line width=0.5pt] (axis cs:8.4,0.41) --  (axis
cs:9.13,0.35) node[above=4.4mm, left=0.95mm] {\scriptsize{100 RAUs, 100 MUs}};

\legend{[right]CVX+SDPT3,[right]CVX+SCS, [right]Matrix Stuffing+SCS}
\end{axis}
\end{tikzpicture}
\caption{The empirical probability of feasibility versus target SINR with different network sizes.}
\label{feasible}
\vspace*{-15pt}
\end{figure}
A unique property of the proposed framework is its infeasibility detection capability, which will be verified in this part. Consider a network with $L=50$ single-antenna RAUs and $K=50$ single-antenna
MUs uniformly and independently
distributed in the square region $[-2000, 2000] \times [-2000, 2000]$
meters.  The empirical probabilities of feasibility in Fig. {\ref{feasible}} show that the proposed framework can detect
the infeasibility accurately compared with the second-order method framework
``{\sf{CVX+SDPT3}}" and the first-order method framework ``{\sf{CVX+SCS}}". Each point
of the simulation
results is averaged over $200$ randomly generated network realizations. The
average (``{\bfseries{total time}}", ``{\bfseries{solving time}}") for obtaining a single point  with ``{\sf{CVX+SDPT3}}", ``{\sf{CVX+SCS}}"
and ``{\sf{Matrix Stuffing+SCS}}" are  ($101.7635$, $99.1140$) seconds, ($5.0754$, $2.3617$) seconds and ($1.8549$, $1.7959$) seconds, respectively. This shows that the operator splitting solver can speedup about 50x over the interior-point solver.

We further consider a larger-sized network with $L=100$ single-antenna RAUs
and $K=100$ single-antenna MUs uniformly and independently
distributed in the square region $[-2000, 2000] \times [-2000, 2000]$
meters. As the second-order method framework fails to scale to this size, we only compare with the first-order method framework. Fig. {\ref{feasible}}
demonstrates that the proposed framework has the same infeasibility detection
capability as the first-order method framework. This verifies
the correctness and the reliability of the proposed fast transformation via matrix stuffing.
Each point of the simulation
results is averaged over 200 randomly generated network realizations. The
average (``{\bfseries{solving time}}", ``{\bfseries{modeling time}}")
for obtaining a single point  with ``{\sf{CVX+SCS}}" and ``{\sf{Matrix Stuffing+SCS}}"
are ($41.9273,18.6079$) seconds and ($31.3660,0.5028$) seconds, respectively. This shows that the matrix stuffing technique can speedup about 40x over the numerical based parser/solver modeling framework CVX. We also note that the solving time of the proposed framework is smaller than the framework ``{\sf{CVX+SCS}}", the speedup is due to the warm-staring \cite[Section 4.2]{Boyd_arXiv2013}.

\subsection{Group Sparse Beamforming for Network Power Minimization}
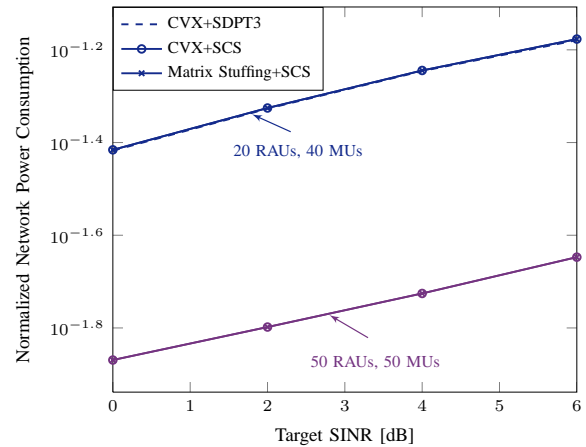
\begin{figure}[t]
\centering
\begin{tikzpicture}[scale=0.9]
\begin{semilogyaxis}[
xmax=6, xmin=0, 
xlabel={Target SINR [dB]}, 
ylabel={Normalized Network Power Consumption}, 
every axis x label/.style={at={(ticklabel cs:0.5)},anchor=near
ticklabel},
every axis y label/.style={at={(ticklabel cs:0.5)},rotate=90, anchor=near
ticklabel}, 
label style={font=\footnotesize},
tick label style={font=\scriptsize},
legend style={at={(0,1)}, anchor=north west, font=\scriptsize}]
\addplot[colorhkust, no marks, dashed, line width=0.8pt] table {totalpower_small_SDPT3.dat};
\addplot[colorhkust, mark=o, mark size=1.6pt, line width=0.8pt]table {totalpower_small_SCS.dat};
\addplot[colorhkust, mark=x, mark size=1.6pt, line width=0.8pt] table {totalpower_small_matscs.dat};

\addplot[colortsinghua, mark=o, mark size=1.6pt, line width=0.8pt] table {totalpower_large_SCS.dat};
\addplot[colortsinghua, mark=x, mark size=1.6pt, line width=0.8pt]table {totalpower_large_matscs.dat};

\draw [-latex', colorhkust, line width=0.5pt] (axis cs:2.3,0.042) --  (axis
cs:1.8,0.0458) node[below =5.5mm,right=-4mm] {\scriptsize{20 RAUs, 40 MUs}};

\draw [-latex', colortsinghua, line width=0.5pt] (axis cs:3.3,0.0151) --  (axis cs:2.8,0.0169) node[below =7mm,right=-4mm] {\scriptsize{50 RAUs, 50 MUs}}; 
\legend{[right]CVX+SDPT3,[right]CVX+SCS, [right]Matrix Stuffing+SCS}
\end{semilogyaxis}
\end{tikzpicture}
\caption{Average normalized network power consumption (i.e., the obtained optimal total network power consumption over the maximum network power consumption with all the RAUs active and full power transmission) versus target SINR with
different network sizes.}
\label{netp}
\vspace*{-15pt}
\end{figure}

In this part, we simulate the network power minimization problem using the group sparse
beamforming algorithm \cite[Algorithm 2]{Yuanming_TWC2014}. We set each fronthaul
link power consumption as $5.6 W$ and set the power amplifier efficiency
coefficient for each RAU as $25\%$. In this algorithm, a sequence of convex feasibility problems need to be solved to determine the active RAUs and one convex optimization problem needs to be solved to determine the transmit beamformers. This relies on the infeasibility detection capability of the proposed framework. 

Consider a network with $L = 20$ 2-antenna RAUs and
$K = 40$ single-antenna MUs uniformly and independently
distributed in the square region $[-1000, 1000] \times [-1000, 1000]$
meters. Each point of the simulation results is averaged over 50 randomly
generated network realizations. Fig. {\ref{netp}} demonstrates the
accuracy of the solutions in the network power consumption obtained by the
proposed framework compared with the second-order method framework ``{\sf{CVX+SDPT3}}"
and the first-order method framework ``{\sf{CVX+SCS}}". The
average (``{\bfseries{total time}}", ``{\bfseries{solving time}}")
for obtaining a single point  with ``{\sf{CVX+SDPT3}}", ``{\sf{CVX+SCS}}"
and ``{\sf{Matrix Stuffing+SCS}}" are  ($48.6916$, $41.0316$) seconds, ($9.4619$,
$1.7433$) seconds and ($2.4673$, $2.3061$)
seconds, respectively. This shows
that the operator splitting solver can speedup about 20x over the interior-point
solver.

We further consider a larger-sized network with $L=50$ 2-antenna RAUs  and
$K=50$ single-antenna MUs  uniformly and independently
distributed in the square region $[-3000, 3000] \times [-3000, 3000]$
meters. As the second-order method framework is not applicable to this problem size, we only compare with the first-order method framework. Each point
of the simulation results is averaged over 50 randomly generated network realizations. Fig. {\ref{netp}}
shows that the proposed framework can achieve the same solutions in network
power consumption as the first-order method framework ``{\sf{CVX+SCS}}". The
average (``{\bfseries{solving time}}", ``{\bfseries{modeling time}}")
for obtaining a single point  with ``{\sf{CVX+SCS}}" and ``{\sf{Matrix Stuffing+SCS}}"
are ($11.9643,69.0520$) seconds and ($14.6559,2.1567$) seconds, respectively. This shows that the matrix stuffing technique can speedup about 30x over
the numerical based parser/solver modeling framework CVX.

In summary,  Fig. {\ref{netp}} demonstrates the capability of infeasibility detection (as a sequence of convex feasibility problems  need to be solved in the RAU selection procedure), the accuracy of the solutions, and speedups provided by the proposed framework over the existing frameworks.

\subsection{Max-min Rate Optimization}
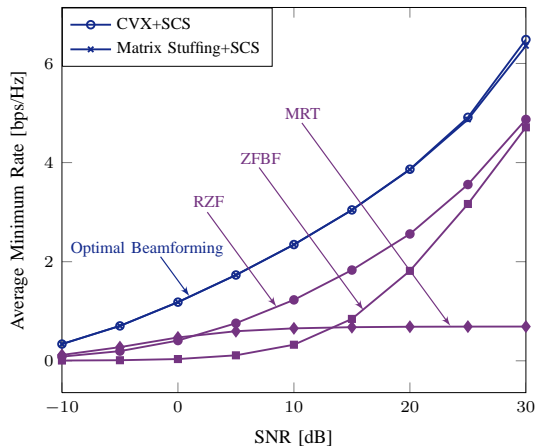
\begin{figure}[t]
\centering
\begin{tikzpicture}[scale=0.9]
\begin{axis}[
xmax=30, xmin=-10, 
xlabel={SNR [dB]}, 
ylabel={Average Minimum Rate [bps/Hz]}, 
every axis x label/.style={at={(ticklabel cs:0.5)},anchor=near
ticklabel},
every axis y label/.style={at={(ticklabel cs:0.5)},rotate=90, anchor=near
ticklabel}, 
label style={font=\footnotesize},
tick label style={font=\scriptsize},
legend style={at={(0,1)}, anchor=north west, font=\scriptsize}]
\addplot[colorhkust, mark=o, mark size=1.6pt, line width=0.8pt] table {Maxmin_opt_SCS.dat};
\addplot[colorhkust,mark=x, mark size=1.6pt, line width=0.8pt] table {Maxmin_opt_matscs.dat};
\addplot[colortsinghua, mark=*, mark size=1.6pt, line width=0.8pt] table {Maxmin_RZF_SDPT3.dat};
\addplot[colortsinghua, mark=square*, mark size=1.3pt, line width=0.8pt]table {Maxmin_ZFBF_SDPT3.dat};
\addplot[colortsinghua, mark=diamond*, mark size=2.1pt, line width=0.8pt] table {Maxmin_MRT_SDPT3.dat};

\draw [-latex', colorhkust, line width=0.5pt] (axis cs:-1.5,2.1) --  (axis
cs:1,1.3) node[above=7mm, left=-6.3mm] {\scriptsize{Optimal Beamforming}};

\draw [-latex', colortsinghua, line width=0.5pt] (axis cs:2.5313,3) --  (axis
cs:8.5,1.09) node[above=15.5mm, left=6.5mm] {\scriptsize{RZF}};

\draw [-latex', colortsinghua, line width=0.5pt] (axis cs:7,3.9) --  (axis
cs:16,1.02) node[above=22.5mm, left=11mm] {\scriptsize{ZFBF}};

\draw [-latex', colortsinghua, line width=0.5pt] (axis cs:10.6875,4.8) --  (axis
cs:23.5,0.7) node[above=31.5mm, left=18mm] {\scriptsize{MRT}};
  
\legend{[right]CVX+SCS,  [right]Matrix Stuffing+SCS}
\end{axis}
\end{tikzpicture}
\caption{The minimum network-wide achievable versus transmit SNR with 55
single-antenna RAUs and 50 single-antenna MUs.}
\label{maxmin_snr}
\vspace*{-15pt}
\end{figure}

We will simulate the  minimum network-wide achievable rate maximization problem
using the max-min fairness optimization algorithm in \cite[Algorithm 1]{Yuanming_Globecom2014} via the bi-section method, which requires to solve a sequence of convex feasibility problems. 
We will not only  show the quality of the solutions and speedups provided by the proposed framework, but also demonstrate that the optimal coordinated beamformers significantly outperform the low-complexity and heuristic transmission strategies, i.e., zero-forcing beamforming (ZFBF)  \cite{Zhang2010adaptive,Goldsmith_JSAC2006optimality}, regularized zero-forcing beamforming (RZF) \cite{Peel_TC2005vector} and maximum ratio transmission (MRT) \cite{Rusek_SPM2013}.  
 
Consider a network with $L = 55$ single-antenna RAUs and
$K = 50$ single-antenna MUs uniformly and independently
distributed in the square region $[-5000, 5000] \times [-5000, 5000]$
meters. Fig. {\ref{maxmin_snr}} demonstrates the minimum network-wide achievable
rate versus different SNRs (which is defined as the transmit power at all the RAUs over the receive noise power at all the MUs) using different algorithms. Each point of the
simulation results is averaged over 50 randomly generated network realizations.
For the optimal beamforming, this figure shows the accuracy of the solutions
obtained by the proposed framework compared with the first-order method framework ``{\sf{CVX+SCS}}". The
average  (``{\bfseries{solving time}}", ``{\bfseries{modeling time}}")
for obtaining a single point  for the optimal beamforming with 
``{\sf{CVX+SCS}}" and ``{\sf{Matrix Stuffing+SCS}}"
are  ($176.3410$, $55.1542$) seconds and ($82.0180$, $1.2012$)
seconds, respectively. This shows that the proposed framework can reduce both the solving time and modelling time via warm-starting and matrix stuffing, respectively.  

Furthermore, this figure also shows that the optimal
 beamforming can achieve quite an improvement for the per-user rate compared
to suboptimal transmission strategies RZF, ZFBF and MRT, which clearly shows the importance of developing optimal beamforming algorithms for such networks. The average (``{\bfseries{solving time}}", ``{\bfseries{modeling time}}")
 for a single point using ``{\sf{CVX+SDPT3}}" for the RZF, ZFBF and MRT are  ($2.6210$, $30.2053$) seconds, ($ 2.4592$,
$30.2098$) seconds and ($2.5966
$, $30.2161$) seconds, respectively. Note that the solving time is very small, which is because we only need to solve a sequence of linear programming problems for power control when the directions of the beamformers are fixed during the bi-section search procedure. The main time consuming part is from transformation using CVX.

\section{Conclusions and Further Works}
\label{concl}
In this paper, we proposed a unified two-stage  framework for large-scale optimization in dense wireless cooperative networks. We showed that various performance optimization problems can be essentially solved by solving one or a sequence of convex optimization or feasibility problems. The proposed framework only requires the convexity of the underlying problems (or subproblems) without any other structural assumptions, e.g., smooth or separable functions. This is achieved by first transforming the original convex problem to the standard form via matrix stuffing and then using the ADMM algorithm to solve the homogeneous self-dual embedding of the primal-dual pair of the transformed standard cone program. Simulation results demonstrated the infeasibility detection capability, the modeling flexibility and computing scalability, and the reliability  of the proposed framework.   

 In principle, one may apply the proposed framework to any large-scale convex optimization problems and only needs to focus on the standard form reformulation as shown in Appendix, as well as to compute the proximal operators for different cone projections in (\ref{proxop}). However, in practice, we need to address the following issues to provide a user-friendly framework and to assist practical implementation:
\begin{itemize}
\item Although the parse/solver frameworks like CVX can automatically transform an original convex problem into the standard form \emph{numerically} based on the graph implementation, extending such an idea to the \emph{automatic and symbolic} transformation, thereby enabling matrix stuffing, is desirable but challenging in terms of reliability  and correctness verification.

\item Efficient projection algorithms are highly desirable. For the subspace projection, as discussed in Section {\ref{cfos}}, parallel factorization and symbolic factorization are especially suitable for the cloud computing environments as in Cloud-RAN \cite{mobile2011c, Yuanming_TWC2014}. For the cone projection, although the projection on the second-order cone is very efficient, as shown in (\ref{proxop}), projecting on the semidefinite cone (which is required to solve the semidefinite programming  problems) is computationally expensive, as it requires to perform eigenvalue
decomposition \cite{Yuanming_ICC2015SDP}. The structure of the cone projection should be exploited to make speedups.

\item It is interesting to apply the proposed framework to various non-convex optimization problems. For instance, the well-known majorization-minimization optimization provides a principled way to solve the general non-convex problems, whereas a sequence of convex subproblems need to be solved at each iteration. Enabling scalable computation at each iteration will hopefully lead to scalability of the overall algorithm.      
\end{itemize}

\section*{Appendix\\
Conic Formulation for  Convex Programs}
We shall present a systematic way to transform the original
problem to the standard convex cone programming form.  We first take
the real-field problem $\mathscr{P}$ with the objective function $f({\bf{x}})=\|{\bf{v}}\|_2$
as an example. At the end of this subsection, we will show how to extend it to the complex-field. 

According to the principle of the disciplined convex programming \cite{boyd2008graph}, the original
problem $\mathscr{P}$  can be rewritten as the following disciplined convex
programming form \cite{boyd2008graph}
\begin{eqnarray}
\!\!\!\!\!\!\!\!\!\!\!\mathscr{P}_{\textrm{cvx}}:\mathop {\rm{minimize~}} && \|{\bf{v}}\|_2\nonumber\\
\label{cvx1}
{\rm{subject~to~}}&& \|{\bf{D}}_l{\bf{v}}\|_2\le
\sqrt{P_l}, l=1,\dots, L\\
\label{cvx2}
&&\|{\bf{C}}_k{\bf{v}}+{\bf{g}}_k\|_2\le
\beta_{k}{\bf{r}}_k^{T}{\bf{v}}, k=1,\dots, K,
\end{eqnarray}
where ${\bf{D}}_{l}={\rm{blkdiag}}\{{\bf{D}}_l^1,\dots,{\bf{D}}_l^K\}\in\mathbb{R}^{N_l
K\times {NK}}$ with ${\bf{D}}_{l}^k=\left[{\bf{0}}_{N_{l}\times\sum_{i=1}^{l-1}N_i},{\bf{I}}_{N_l\times
N_l}, {\bf{0}}_{{N_l}\times
\sum_{i=l+1}^{L}N_i}\right]\in\mathbb{R}^{N_{l}\times N}$, $\beta_k=\sqrt{1+1/\gamma_k}$,  ${\bf{r}}_k=\left[{\bf{0}}_{(k-1)N}^{T},{\bf{h}}_k^{T},
{\bf{0}}_{(K-k)N}^{T}\right]^{T}\in\mathbb{R}^{NK}$, ${\bf{g}}_k=[{\bf{0}}_K^T,
\sigma_k]^T\in\mathbb{R}^{K+1}$, and ${\bf{C}}_k=[\tilde{\bf{C}}_k, {\bf{0}}_{NK}]^{T}\in\mathbb{R}^{(K+1)\times
NK}$ with $\tilde{\bf{C}}_k={\rm{blkdiag}}\{{\bf{h}}_k, \dots, {\bf{h}}_k\}\in\mathbb{R}^{NK\times K}$. It is thus easy to check the convexity of problem $\mathscr{P}_{\textrm{cvx}}$, following the disciplined convex programming
ruleset \cite{boyd2008graph}.

\subsection{Smith Form Reformulation}
To arrive at the standard convex cone program $\mathscr{P}_{\textrm{cone}}$,
we rewrite problem $\mathscr{P}_{\textrm{cvx}}$ as the following Smith form \cite{smith1996optimal} by
introducing a new variable for each subexpression in $\mathscr{P}_{\textrm{cvx}}$,
\begin{eqnarray}
\label{smithr}
\mathop {\rm{minimize~}}&&  x_{0}\nonumber\\
{\rm{subject~to~}}&& \|{\bf{x}}_1\|=x_0, {\bf{x}}_1={\bf{v}} \nonumber\\
&&\mathcal{G}_1(l), \mathcal{G}_2(k),\forall k, l,
\end{eqnarray}
where $\mathcal{G}_1(l)$ is the Smith form reformulation for the transmit
power constraint for RAU $l$ (\ref{cvx1}) as follows  
\begin{eqnarray}
\mathcal{G}_1(l):\left\{\begin{array}{l}
(y_0^l, {\bf{y}}_1^l)\in\mathcal{Q}^{KN_l+1}\\
y_0^{l}=\sqrt{P_{l}}\in\mathbb{R}\\
{\bf{y}}_1^l={\bf{D}}_l{\bf{v}}\in\mathbb{R}^{KN_{l}},
\end{array}\right.
\end{eqnarray}
and $\mathcal{G}_2(k)$ is the Smith form reformulation for the QoS constraint
for MU $k$ (\ref{cvx2}) as follows
\begin{eqnarray}
\mathcal{G}_2(k):\left\{\begin{array}{l}
(t_0^k, {\bf{t}}_1^k)\in\mathcal{Q}^{K+1}\\
t_0^k=\beta_{k}{\bf{r}}_k^{T}{\bf{v}}\in\mathbb{R}\\
{\bf{t}}_1^k={\bf{t}}_2^k+{\bf{t}}_3^k\in\mathbb{R}^{K+1}\\
{\bf{t}}_2^k={\bf{C}}_k{\bf{v}}\in\mathbb{R}^{K+1}\\
{\bf{t}}_3^k={\bf{g}}_k\in\mathbb{R}^{K+1}.
\end{array}\right.
\end{eqnarray}

Nevertheless, the Smith form reformulation (\ref{smithr}) is not convex due
to the non-convex constraint $\|{\bf{x}}_1\|=x_0$. We thus relax the non-convex
constraint as $\|{\bf{x}}_1\|\le x_0$, yielding the following relaxed Smith
form 
\begin{eqnarray}
\label{rsm}
\mathop {\rm{minimize~}} && x_{0}\nonumber\\
{\rm{subject~to~}}&& \mathcal{G}_{0}, \mathcal{G}_1(l), \mathcal{G}_2(k),\forall k, l, \end{eqnarray}
where 
\begin{eqnarray}
\mathcal{G}_0:\left\{\begin{array}{l}
(x_0, {\bf{x}}_1)\in\mathcal{Q}^{NK+1}\\
{\bf{x}}_1={\bf{v}}\in\mathbb{R}^{NK}.
\end{array}\right.
\end{eqnarray}
It can be easily proved that the constraint $\|{\bf{x}}_1\|\le x_0$
has to be active at the optimal solution; otherwise, we can always scale
down $x_0$ such that the cost function can be further minimized while still
satisfying the constraints. Therefore, we conclude that the relaxed Smith
form (\ref{rsm}) is equivalent to the original problem $\mathscr{P}_{\textrm{cvx}}$.

\subsection{Conic Reformulation}
Now, the relaxed Smith form reformulation (\ref{rsm}) is readily to be reformulated
as the standard cone programming form $\mathscr{P}_{\textrm{cone}}$. Specifically,
define the optimization variables $[x_0; {\bf{v}}]$ with the same order
of equations
as in $\mathcal{G}_0$, then $\mathcal{G}_0$ can be rewritten as 
\begin{eqnarray}
{\bf{M}}[{{x}}_0; {\bf{v}}]+{\bs{\mu}}_0={\bf{m}},
\end{eqnarray}
where the slack variables belong to the following convex set 
\begin{eqnarray}
{\bs{\mu}}_0\in\mathcal{Q}^{NK+1},
\end{eqnarray}
and ${\bf{M}}\in\mathbb{R}^{(NK+1)\times(NK+1)}$
and ${\bf{m}}\in\mathbb{R}^{NK+1}$ are given as follows
\begin{eqnarray}
{\mathbf{M}} =
\left[ \begin{array}{r|c}
-1&   \\
\hline
&   -{\bf{I}}_{NK}
 
\end{array} \right],
{{\bf{m}}}=\left[ \begin{array}{c}
0\\
\hline
{\bf{0}}_{NK}
\end{array} \right],
\end{eqnarray}
respectively. Define the optimization variables $[y_0^l; {\bf{v}}]$ with the same order
of equations as in $\mathcal{G}_1(l)$, then $\mathcal{G}_1(l)$ can be rewritten
as
\begin{eqnarray}
{\bf{P}}^l[{{y}}_0^{l}; {\bf{v}}]+{\bs{\mu}}_1^l={\bf{p}}^l,
\end{eqnarray}
where the slack variables ${\bs{\mu}}_1^l\in\mathbb{R}^{KN_l+2}$ belongs
to the following convex set formed by the Cartesian product of two convex
sets\begin{eqnarray}
{\bs{\mu}}_1^l\in\mathcal{Q}^{1}\times\mathcal{Q}^{KN_l+1},
\end{eqnarray}
and ${\bf{P}}^l\in\mathbb{R}^{(KN_l+2)\times (NK+1)}$ and ${\bf{p}}^{l}\in\mathbb{R}^{KN_l+2}$
are given as follows 
\begin{eqnarray}
{\mathbf{P}}^l =
\left[ \begin{array}{r|c}
1&  \\
\hline
-1&   \\
&   -{\bf{D}}_l
 
\end{array} \right],
{{\bf{p}}}^l=\left[ \begin{array}{c}
\sqrt{P_l} \\
\hline
0\\
{\bf{0}}_{KN_l}
\end{array} \right],
\end{eqnarray}
respectively. Define the optimization variables $[t_0^k; {\bf{v}}]$ with the same order
of equations as in $\mathcal{G}_2(k)$, then $\mathcal{G}_2(k)$ can be rewritten
as 
\begin{eqnarray}
{\bf{Q}}^k[t_{0}^k;{\bf{v}}]+{\bs{\mu}}_2^k={\bf{q}}^k,
\end{eqnarray}
where the slack variables ${\bs{\mu}}_2^k\in\mathbb{R}^{K+3}$ belong to the
following convex set formed by the Cartesian product of two convex sets
\begin{eqnarray}
{\bs{\mu}}_2^k\in\mathcal{Q}^{1}\times \mathcal{Q}^{K+2},
\end{eqnarray}
and ${\bf{Q}}^k\in\mathbb{R}^{(K+3)\times (NK+1)}$ and ${\bf{q}}^{k}\in\mathbb{R}^{K+3}$
are given as follows
\begin{eqnarray}
{\mathbf{Q}}^k =
\left[ \begin{array}{r|c}
1& -\beta_{k}{\bf{r}}_k^{T} \\
\hline
-1&   \\
&   -{\bf{C}}_k
 
\end{array} \right], 
{{\bf{q}}}^k=\left[ \begin{array}{c}
0 \\
\hline
0\\
{\bf{g}}_k
\end{array} \right],
\end{eqnarray}
respectively.

Therefore, we arrive at the standard form $\mathscr{P}_{\textrm{cone}}$ by
writing the optimization variables ${\bs{\nu}}\in\mathbb{R}^{n}$ as follows
\begin{eqnarray}
\label{solv}
{\bs{\nu}}=[x_0; y_0^{1};\dots; y_0^{L}; t_0^1; \dots, t_0^K; {\bf{v}}],
\end{eqnarray}
and ${\bf{c}}=[1; {\bf{0}}_{n-1}]$. The structure of the standard cone programming
$\mathscr{P}_{\textrm{cone}}$
 is characterized by the following data
\begin{eqnarray}
\label{ds1}
n&=&1+L+K+NK,\\
m&=&(L+K)+(NK+1)+\sum_{l=1}^{L}(KN_l+1)\!+\!K(K+2),\\
\mathcal{K}&=& \underbrace{\mathcal{Q}^{1}\times\cdots\times\mathcal{Q}^1}_{L+K}\times\mathcal{Q}^{NK+1}\times\underbrace{\mathcal{Q}^{KN_1+1}\times\dots\times\mathcal{Q}^{KN_L+1}}_{L}\times\nonumber\\
&&\underbrace{\mathcal{Q}^{K+2}\times\dots\times\mathcal{Q}^{K+2}}_{K},
\end{eqnarray}
where $\mathcal{K}$ is the Cartesian product of $2(L+K)+1$ second-order 
ones, and
 ${\bf{A}}$ and $\bf{b}$ are given as
follows: 
\begin{eqnarray}
\label{ds2}
\!\!\!\!\!\!\!\!{\mathbf{A}} =
\left[ \begin{array}{c|ccc|ccc|c}
&1&&&&&\\
&&\ddots&&&&\\
&&&1&&&\\
\hline
&&&&1&&&-\beta_{1}{\bf{r}}_1^{T}\\
&&&&&\ddots&&\vdots\\
&&&&&&1&-\beta_{K}{\bf{r}}_K^{T}\\
\hline
-1&&&&&&&\\
&&&&&&&-{\bf{I}}_{NK}\\
\hline
&-1&&&&&&\\
&&&&&&&-{\bf{D}}_1\\
\hline
&&\vdots&&&\vdots&&\vdots\\
\hline
&&&-1&&&&\\
&&&&&&&-{\bf{D}}_L\\
\hline
&&&&-1&&&\\
&&&&&&&-{\bf{C}}_1\\
\hline
&&&&&\vdots&&\vdots\\
\hline
&&&&&&-1&\\
&&&&&&&-{\bf{C}}_K\\
\end{array} \right],\!
{\bf{b}}=\left[ \begin{array}{c}
\sqrt{P_1}\\
\vdots\\
\sqrt{P_L}\\
\hline
0\\
\vdots\\
0\\
\hline
0\\
{\bf{0}}_{NK}\\
\hline
0\\
{\bf{0}}_{KN_1}\\
\hline
\vdots\\
\hline
0\\
{\bf{0}}_{KN_1}\\
\hline
0\\
{\bf{g}}_1\\
\hline
\vdots\\
\hline
0\\
{\bf{g}}_K
\end{array}\right],\end{eqnarray}
respectively.

\subsection{Matrix Stuffing}
Given a problem instance $\mathscr{P}$, to arrive at the standard cone program
form, we only need to copy the parameters
of the maximum transmit power $P_{l}$'s to the data of the 
standard form, i.e.,  $\sqrt{P_l}$'s in $\bf{b}$, copy the parameters of
the SINR
thresholds ${\bs{\gamma}}$ to the data of the standard form, i.e., $\beta_k$'s
in ${\bf{A}}$, and copy the parameters of the channel realizations ${\bf{h}}_k$'s
to the data of the standard form, i.e., ${\bf{r}}_k$'s and ${\bf{C}}_k$'s
in $\bf{A}$. As we only need to perform copying the memory for the transformation,
this procedure can be very efficient compared to the state-of-the-art numerical
based modeling
frameworks like CVX. 

\subsection{Extension to the Complex Case}
\label{complex}
 For ${\bf{h}}_k\in\mathbb{C}^{N},
{\bf{v}}_{i}\in\mathbb{C}^{N}$, we have
\begin{eqnarray}
{\bf{h}}_k^{\sf{H}}{\bf{v}}_i\Longrightarrow
{\underbrace{\left[ \begin{array}{cc}
\mathfrak{R}({\bf{h}}_k)\ & -\mathfrak{J}({\bf{h}}_k) \\
\mathfrak{J}({\bf{h}}_k)\ & \mathfrak{R}({\bf{h}}_k) 
\end{array} \right]}_{\tilde{\bf{h}}_k}}^{T}\underbrace{\left[ \begin{array}{c}
\mathfrak{R}({\bf{v}}_i)\ \\
\mathfrak{J}({\bf{v}}_i) 
\end{array} \right]}_{\tilde{\bf{v}}_i},
\end{eqnarray} 
where $\tilde{\bf{h}}_k\in\mathbb{R}^{2N\times
2}$ and $\tilde{\bf{v}}_i\in\mathbb{R}^{2N}$. Therefore, the complex-field
problem can be changed into the real-field problem by
the transformations: ${\bf{h}}_k\Rightarrow\tilde{\bf{h}}_k~{\textrm{and}}~{\bf{v}}_i\Rightarrow\tilde{\bf{v}}_i$.

\section*{Acknowledgment}

The authors would like to thank Dr. Eric Chu for his insightful discussions and constructive comments related to this work.

\bibliographystyle{ieeetr}

\end{document}